\documentclass[12pt]{article}
\usepackage{amsmath}
\usepackage{graphicx,psfrag,epsf}
\usepackage{enumerate}
\usepackage{url}

\newcommand{\blind}{0}

\usepackage{amsmath}
\usepackage{graphicx,psfrag,epsf}
\usepackage{enumerate}

\usepackage{amssymb}
\usepackage{soul}
\usepackage{amsmath}
\usepackage{amsthm}
\usepackage{latexsym}
\usepackage{verbatim}
\usepackage{epsfig}

\usepackage{amsbsy}
\usepackage{amscd}
\usepackage{bm}
\usepackage{graphicx,psfrag,epsf}
\usepackage{enumerate}
\usepackage[authoryear,sort]{natbib}
\usepackage{bm}
\usepackage{color}
\usepackage{nicefrac}

\providecommand{\M}[1]{\mathbf#1}
\providecommand{\mc}[1]{\mathcal#1}
\providecommand{\mc}[1]{\mathcal#1}

\newcommand{\R}{{\mathbb R}}

\DeclareMathOperator{\E}{\mathbf{E}}
\DeclareMathOperator{\p}{\mathbf{P}}

\DeclareMathOperator{\var}{Var}
\DeclareMathOperator{\cov}{Cov}
\newcommand{\indep}{\rotatebox[origin=c]{90}{$\models$}}

\DeclareMathOperator{\tr}{tr}
\DeclareMathOperator{\diag}{diag}
\providecommand{\T}{\top} 

\DeclareMathOperator*{\argmin}{argmin}

\providecommand{\wt}[1]{\widetilde{#1}}
\providecommand{\wh}[1]{\widehat{#1}}

\providecommand{\nnorm}[1]{ \lVert#1 \rVert}
\newcommand{\scp}[2]{\left\langle#1, #2\right\rangle}
\newcommand{\nscp}[2]{\langle#1, #2\rangle}

\newcommand{\blanco}[1]{  }

\newcommand{\deriv}[3]{%
\ifthenelse{#1 = 1}{\frac{d\,#2}{d\,#3}}{\frac{d^{{#1}} #2}{d{#3}^{{#1}}}}
}

\newcommand{\partials}[3]{%
\ifthenelse{#1 = 1}{\frac{\partial\,#2}{\partial\,#3}}{\frac{\partial^{#1}
    #2}{\partial#3^{#1}}}
}

\def \coloneq{\mathrel{\mathop:}=}
\def \invcoloneq{=\mathrel{\mathop:}}

\begin{document}

\def\spacingset#1{\renewcommand{\baselinestretch}%
{#1}\small\normalsize} \spacingset{1}

\newcommand{\circled}[1]{\small{\raisebox{.6pt}{\textcircled{\raisebox{-.8pt}{#1}}}}}

\if0\blind
{
  \title{\bf Accounting for Mismatch Error in Small Area Estimation with Linked Data}
  \author{Enrico Fabrizi$^{1}$ $\qquad$ Nicola Salvati$^2{}$ $\qquad$ Martin Slawski$^3$\thanks{Partially supported by NSF grant \#2120318}\\[2ex]
 $^{1}${\normalsize DISES \& DSS, Universit\`a Cattolica del S. Cuore, Italy}\\ 
 $^{2}${\normalsize Department of Economics and Management, University of Pisa, Italy} \\
 $^3${\normalsize Department of Statistics, George Mason University, USA}}
  \maketitle
} \fi

\if1\blind
{
  \bigskip
  \bigskip
  \bigskip
  \begin{center}
    {\LARGE\bf Title}
\end{center}
  \medskip
} \fi

\bigskip
\begin{abstract}
In small area estimation different data sources are integrated in order to produce reliable estimates of target parameters (e.g.,~a mean or a proportion) for a collection of small subsets (areas) of a finite population. Regression models such as the linear mixed effects model or M-quantile regression are often used to improve the precision of survey sample estimates by leveraging auxiliary information for which means or totals are known at the area level. In many applications, the unit-level linkage of records from different sources is probabilistic and potentially error-prone. In this paper, we present adjustments of the small area predictors that are based on either the linear mixed effects model or M-quantile regression to account for the presence of linkage error. These adjustments are developed from a two-component mixture model that hinges on the assumption of independence of the target and auxiliary variable given incorrect linkage. Estimation and inference is based on composite likelihoods and machinery revolving around the Expectation-Maximization Algorithm. For each of the two regression methods, we propose modified small area predictors and approximations for their mean squared errors. The empirical performance of the proposed approaches is studied in both design-based and model-based simulations that include comparisons to a variety of baselines.       
\end{abstract}


\spacingset{1.45}
\section{Introduction}
\label{sec:intro}
Estimates of finite population parameters such as means or totals that can be obtained from sample surveys may be needed for the population as a whole or for a collection of subsets known as domains or areas, although they are not necessarily geographical regions.
A small area estimation problem arises when the area-specific sample sizes are such that estimates relying only on observations from that area have inadequate precision in most or all cases. For instance, this can happen when the estimates are needed for small administrative subdivisions of a country. 

Small area estimation methodology is about the integration of a survey and other sources to improve the estimation of area-specific parameters. Recent reviews of the small area literature can be found, e.g., in \cite{rao2015, pfeffermann2013new, sugasawa2020small}. According to the terminology used in this rich literature, in the present paper we focus on unit-level methods relying on linear models \citep[][chapter 7]{rao2015}. In this context a classical scenario is that in addition to the variable of interest a number of auxiliary variables are recorded in the same survey, and means or totals for these auxiliary variables are accurately known from some external source. A model relating the target and the auxiliary variables can therefore be used for improving small area estimates. 

In recent years, it has become more common that unit-level values of the auxiliary variables, not only their totals, are from a different data source that is linked at the individual level with the survey sample. Data integration has gained in popularity as information technology makes linkage of the records pertaining to the same individual stored in many registers possible \citep[see for instance][for general reviews]{shlomo2014probabilistic, sayers2016probabilistic}. On the one hand, this provides significant potential for enhancing small area estimation methods. 
On the other hand, the linkage of different sources at the micro-level is rarely error-free. Measurement error can be an issue \citep{boudreaux2015measurement}; moreover, missing or false matches can occur whenever unique identifiers are missing or corrupted for a fraction of the cases. Record linkage involve delicate data disclosure issues so that the agent carrying out the linkage and the data analyst operating on the linked data are usually separate individuals. In this paper we will adopt the secondary analyst point of view, assuming that the researcher who does the small area estimation has limited or no information on the linkage process. We will not discuss other potential sources of measurement error and instead focus on the impact of mismatches. 

The literature on the secondary analysis of linked data is focused primarily on linear regression, where mismatches are known to cause bias and standard error inflation in regression slope estimates \citep{scheuren93}. A first stream of contributions \citep{Scheuren97, Lahiri05, Chambers2009, di2018adjusting, Han2019, Chambers2019improved,zhang2021linkage} introduce assumptions on the linkage process and focus on bias correction. A second stream of literature noting that the impact of mismatched records is similar to that of (not representative) outliers introduce the use of robust regression with very weak assumptions on the linkage process \citep{SlawskiBenDavid2017, slawski2021pseudo, chambers2023robust, wang2022regression}. The study of linear {\em mixed} models with linked data is much more limited \citep{samart2014linear, han2018statistical} and so it is for small area estimation \citep{briscolini2018new, salvati2021}.

We focus on small area methods under the assumption that auxiliary and target variables are from different sources, the linkage between the two sources is probabilistic and linkage errors are possible. Similarly to \cite{salvati2021} we focus on small area predictors based on linear mixed models (LMM) and M-quantile (MQ) regression. Different from \cite{salvati2021} our assumptions on the information available on the linkage process are weaker, following \cite{slawski2021pseudo, Slawski2023}. Specifically, we only assume that the fraction of wrongly linked units in the sample is bounded but no additional information on the linkage errors associated with either individual units or blocks of units is required. At the same time, such information as well as other variables indicative of the match status of the linked records can be readily incorporated.  

We represent regression models in the presence of linkage errors as a mixture of two components assuming the independence of target and auxiliary variable given a wrong link has taken place. This approach, which has been proposed by \cite{slawski2021pseudo} for linear regression, is thereby extended to linear mixed models and M-quantile regression. Regarding estimation, we consider an approach based on composite likelihood estimation \citep{Lindsay1988, Varin2011} whose implementation relies on the Expectation-Maximization (EM) algorithm \citep{Dempster1977}.

In summary, we contribute to the literature by proposing a set of new small area predictors constructed from linked data contaminated by mismatch error under weak assumptions regarding the available information on the linkage process. We study the properties of the proposed predictors and develop estimators for their mean squared errors. Our empirical evaluation demonstrates the effectiveness of our approach, achieving a performance that is at least on par with the methods in \cite{salvati2021} while requiring less information. 

The paper is organized as follows. In Section \ref{sec:meth}, we introduce notation, the assumptions on the linkage process and review linear mixed models and linear M-quantile regression methods on which our small area predictors are based. The predictor of area-level means based on linear mixed models is introduced in Section \ref{LMM_predictor} along with the EM algorithm used for estimation and the proposed mean square error estimator. The same is done for the predictor based on M-quantile regression in Section \ref{MQ_predictor}. Section \ref{simulations} contains an empirical evaluation of the proposed predictors. Specifically, we consider two alternative simulation studies. The first is model-based in the sense that a synthetic population is generated under a linear mixed model with different assumptions on the distribution of random effects and random linkage errors. The second is design-based in that a real finite population is considered in which only sampling and linkage are simulated. The simulation studies are also used to evaluate the empirical properties of the proposed mean square error estimators. Finally, Section \ref{conclusions} contains a discussion of the main results and conclusions. A variety of technical details have been deferred to an appendix.


\section{Settings} \label{sec:meth}
Let us consider a finite population of size $N$ partitioned into $D$ domains of size $N_j$, so that $N = \sum_{j = 1}^D N_j$. We will refer to these domains as \emph{small areas} (SA) in line with the discussion in the introduction.

Each unit in the population is associated with an outcome variable of interest $y_{ij}$, $1 \leq i \leq N_j$, $1 \leq j \leq D$. The collections of all outcomes per small area $j$ can be arranged into vectors $\M{y}_{j} = (y_{ij})_{i = 1}^{N_j}$, $1 \leq j \leq D$. Accordingly, we let $\M{y} = [\M{y}_{1}^{\T} \, \ldots \, \M{y}_{D}^{\T}]^{\T}$ denote the vertical concatenation of these vectors. 
We assume that the main target of inference is to estimate the collection of SA means 
\[
\overline{y}_{j} = \frac{1}{N_j} \sum_{i = 1}^{N_j} y_{ij}, \quad 1 \leq j \leq D.
\]
We assume that for each unit in the population, there exist pairs of covariates $\{ (\M{x}_{ij}, \M{z}_{ij}) \}_{i = 1, \; j=1}^{N_j, \; \; D}$ taking values in $\R^p$ and $\R^q$, respectively. Moreover, let $\M{X}_{j}$ and $\M{Z}_{j}$ denote the 
   design matrices with rows $\{ \M{x}_{ij}^{\T} \}_{i = 1}^{N_j}$ and $\{ \M{z}_{ij}^{\T} \}_{i = 1}^{N_j}$, respectively, $1 \leq j \leq D$,
   and let $\M{X}$ and $\M{Z}$, respectively, denote the vertical concatenation of these matrices. 

\subsection{Assumptions about sampling and linkage}\label{assumptions}
The \emph{latent} correct linkage of the outcome $\M{y}$ and the covariates $(\M{X}, \M{Z})$ can be represented by the column-concatenated matrix $\bm{\Omega} = [\M{y} \;\, \M{X} \,\, \M{Z}]$. Linkage at the population level
in the presence of linkage error is assumed to yield the matrix $\bm{\Omega}^{\star } = [(\M{\Pi}^{\star} \M{y}) \;\, \M{X} \,\, \M{Z}]$, where $\bm{\Pi}^{\star}$ is a latent random $N$-by-$N$ permutation matrix. In other words, it is assumed that $\M{y}$ and $(\M{X}, \M{Z})$ are from registers that contain no duplicates, correspond to the same target population and can in principle be associated on a one-to-one basis (complete linkage).
This assumption can be relaxed to allow for incomplete linkage as long as the occurrence of missing links is non-informative given $\M{X}$. More importantly, we assume that linkage error is independent of the outcomes and covariates, i.e., $\bm{\Pi}^{\star} \indep \,   \{ \M{y}, \M{X}, \M{Z} \}$.

The secondary data analyst is assumed to have access to a data set of the form  
\[
\bm{\Omega}^{(s)} = \M{S} \bm{\Omega}^{\star } = [ \M{S}\M{\Pi}^{\star} \M{y} \;\; \M{S} \M{X} \;\, \M{S}\M{Z}] = [\M{y}^{(s)} \; \M{X}^{(s)} \; \M{Z}^{(s)}],
\] 
where sampling is encoded by $\M{S}$, a $\{0,1\}$-valued $n \times N$ ($n \leq N$) matrix having a single one per row and column sums at most one.
We let $\M{x}_{ij}^{(s)}$  and $\M{z}_{ij}^{(s)}$, $1 \leq i \leq n_j$, denote the
covariates in the sample of size $n_j$ in area $j$, $1 \leq j \leq D$, $\sum_{j = 1}^D n_j = n$. Accordingly, let $q_{ij} = s_{ij} + \sum_{ k < j} N_k$,  $s_{ij} \in \{1, \ldots, N_j \}$, denote the position of the entry one in the row of $\M{S}$ that extracts $(\M{x}_{ij}^{(s)}, \M{z}_{ij}^{(s)})$ from $[\M{X} \; \M{Z}]$ such that $(\M{x}_{ij}^{(s)}, \M{z}_{ij}^{(s)}) = (\M{x}_{s_{ij}\,j,} \, \M{z}_{s_{ij}\,j})$, $1 \leq i \leq n_j$, $1 \leq j \leq D$; without loss of generality, we assume that $s_{1j} < \ldots < s_{n_j \, j}$, $1 \leq j \leq D$. 

We assume that sampling is non-informative, i.e., $\M{S} \indep \, \M{y} | \{ \M{X}, \M{Z} \}$. We also assume independence of sampling  and linkage, i.e., $\M{S} \indep \bm{\Pi}^{\star}$. 
Eventually, we assume that $\sum_{k \notin \{s_{ij} \}} \M{x}_{kj}$ and $\sum_{k \notin \{s_{ij} \}} \M{z}_{kj}$, $1 \leq j \leq D$, are known and can thereby be used in prediction.

Let $y_{ij}^{(s)}$ denote the outcome observed jointly with 
$(\M{x}_{ij}^{(s)}, \M{z}_{ij}^{(s)})$, $1 \leq i \leq n_j$, $1 \leq j \leq D$. Define mismatch indicators 
$m_{ij}^{(s)} = \mathbb{I}(y_{ij}^{(s)} \neq y_{s_{ij} \, j})$, or equivalently, $m_{ij}^{(s)} = \mathbb{I}(\bm{\Pi}^{\star}_{q_{ij} \, q_{ij}} \neq 1)$. The mismatch indicators are considered latent; the data analyst may only be given additional covariates $\M{d}_{ij}^{(s)}$ predictive of $m_{ij}^{(s)}$, assumed independent of the covariates, i.e., $\M{d}_{ij}^{(s)} \indep \{\M{x}_{ij}^{(s)}, \M{z}_{ij}^{(s)} \}$, $1 \leq i \leq n_j$, $1 \leq j \leq D$. Specifically, we assume that 
\begin{equation*}
\p(m_{ij}^{(s)} = 1 | \M{d}_{ij}^{(s)}) = h \big(\M{d}_{ij}^{(s)}; \bm{\alpha}_{*} \big), \quad 1 \leq i \leq n_j, \; 1 \leq j \leq D,
\end{equation*}
for a known function $h: \R \rightarrow (0,1)$, and an unknown parameter $\bm{\alpha}_*$.  

A key assumption for what follows is that $y_{ij}^{(s)} \indep \; (\M{x}_{ij}^{(s)}, \M{z}_{ij}^{(s)}) | \{ m_{ij}^{(s)} = 1\}$. As a result,
the distribution of $y_{ij}^{(s)}$ given $\{ m_{ij}^{(s)} = 1\}$ is given by the marginal distribution of the sampled $\{ y_{ij} \}$ whose density $g$ is given by:
\begin{equation}\label{eq:non_informative}
g(y) \coloneq f(y | \texttt{s}, \{ m  = 1\}) 
=f(y | \texttt{s}),  
\end{equation}
where $(y, \M{x}, \M{z}, m, \texttt{s})$ refers to a generic element of
$\{ (y_{ij}, \M{x}_{ij}, \M{z}_{ij}, m_{ij}, \texttt{s}_{ij}) \}$ with 
 $\texttt{s}_{ij}$ referring to the event that the unit $i$ in area $j$ is sampled, $1 \leq i \leq N_j$, $1 \leq j \leq D$, and with $m_{ij} = 1$ if the corresponding unit is incorrectly linked, i.e., the corresponding diagonal element of  $\bm\Pi^{\star}$ at position $\sum_{k < j} N_k + i$ is zero, $1 \leq i \leq N_j$, $1 \leq j \leq D $. The equality in Eq.~\eqref{eq:non_informative} follows from the assumption that linkage is independent of sampling and the $y$'s. Expanding $f(y|\texttt{s})$ further, we obtain 
 \begin{align}
g(y) = f(y|\texttt{s}) &= \int f(y|\M{x}, \M{z}, \texttt{s}) \, dP(\M{x}, \M{z}|\texttt{s}) \notag \\
                       &= \int f(y|\M{x}, \M{z}) \, dP(\M{x}, \M{z}|\texttt{s}) \label{eq:marginal_penultimate} \\
                       &= \sum_{j = 1}^D \sum_{i = 1}^{N_j} \; \frac{\p(\texttt{s}_{ij} | \M{x}_{ij}, \M{z}_{ij})}{\sum_{\ell = 1}^{D} \sum_{k = 1}^{N_{\ell}} \p(\texttt{s}_{k\ell} | \M{x}_{k\ell}, \M{z}_{k\ell})} \, \label{eq:marginal_ultimate}
\, f(y|\M{x}_{ij}, \M{z}_{ij}),
 \end{align}
 where $P(\M{x}, \M{z} | \texttt{s})$ denotes the atomic probability measure supported on $\{ (\M{x}_{ij}, \M{z}_{ij}) \}$ with probability masses proportional to the associated sampling probabilities $\{ \p(\texttt{s}_{ij} | \M{x}_{ij}, \M{z}_{ij}) \}$. Furthermore, in Eq.~\eqref{eq:marginal_penultimate} we have invoked the assumption that sampling is non-informative. 
\begin{table}[t!!!]
{\footnotesize
\begin{center}
\begin{tabular}{|ll|ll|}
\hline & & &\\[-3ex]
$D$ & number of areas  & $\nscp{\cdot}{\cdot}$ & Euclidean inner product \\[1ex]
$N_j$ & number of units in area $j$ & $\nnorm{\cdot}_2$  & Euclidean norm \\[1ex]
$n_j$ & number of sampled units in area $j$  & $\mc{P}_j$  & power set of $\{1,\ldots,n_j\}$  \\[1ex] 
$\mc{D}$ & observed data (EM algorithm) & $\ell_j$ & generic element of $\mc{P}_j$ \\
$\M{E}$ & expectation  &  $\lambda_j$ & latent set of correct matches in area $j$ \\[1ex]
$\p$ & probability & $\M{v}_A$  &  sub-vector of vector $\M{v}$ defined by index set $A$ \\[1ex]
$\var$ & variance  &  $\M{M}_A$  & row sub-matrix of matrix $\M{M}$  defined by index set $A$ \\
$\cov$ & covariance &  $^{(s)}$ &  superscript corresponding to sampled units   \\
$\mathbb{I}$ & indicator function &  $^{(t)}$ & iteration counter (EM algorithm)  \\
$I_d$ & identity matrix of dimension $d$  &  $^{[\ldots]}$ & superscript indexing Monte-Carlo samples  \\
$|A|$ & cardinality of set $A$  &  $g$ & (marginal) PDF of sampled $y$'s   \\
$|\M{M}|$ & determinant of matrix $\M{M}$  & $h(\cdot;\bm{\alpha})$ & function s.t.~$\p(m_{ij}^{(s)} | \M{d}_{ij}^{(s)}) = h(\M{d}_{ij}^{(s)};\bm{\alpha})$  \\
$\indep$  & stochastic independence   & $N_d(\bm{\mu}, \Sigma)$ & $d$-dimensional Normal distribution\\
$\varphi$ & standard Normal PDF  &  $\Phi$ & standard Normal CDF      \\
\hline
\end{tabular}
\end{center}}
\vspace*{-3ex}
\caption{Summary of notation used repeatedly in this paper.}\label{tab:notation}
\end{table} 

Note that as far as notation is concerned, the two displays above have made use of the following conventions regarding probability density functions (PDFs), which will be adopted throughout this paper: instead of writing 
$f_{\M{x}}(\M{x}_0)$ for the density of a random vector $\M{x}$ evaluated at a point $\M{x}_0$, we drop the symbol in the subscript and simply write $f(\M{x}_0)$ with the convention that the corresponding random variable is inferred from the symbol in the argument. Similar conventions are adopted for joint and conditional PDFs, i.e, we use $f(\M{a}_0, \ldots, \M{z}_0)$ instead of $f_{\M{a} \ldots \M{z}}(\M{a}_0, \ldots, \M{z}_0)$ and $f(\M{x}_0|\M{y}_0)$ instead of $f_{\M{x}|\M{y} = \M{y}_0}(\M{x}_0)$, etc. Subscripts in $f$ will be present in case there is no argument. By default, symbols will be boldfaced to indicate vectors or matrices. The dependence of PDFs on parameters is expressed via $f(\cdot\,;\ldots)$, where $\ldots$ represents a list of parameters. For the convenience of the reader, notation that is used frequently in this paper is summarized in Table \ref{tab:notation}.

\subsection{The linear mixed effects SA model}\label{saemodel_lmm}
\begin{figure}
\begin{center}
\includegraphics[width = 0.6\textwidth]{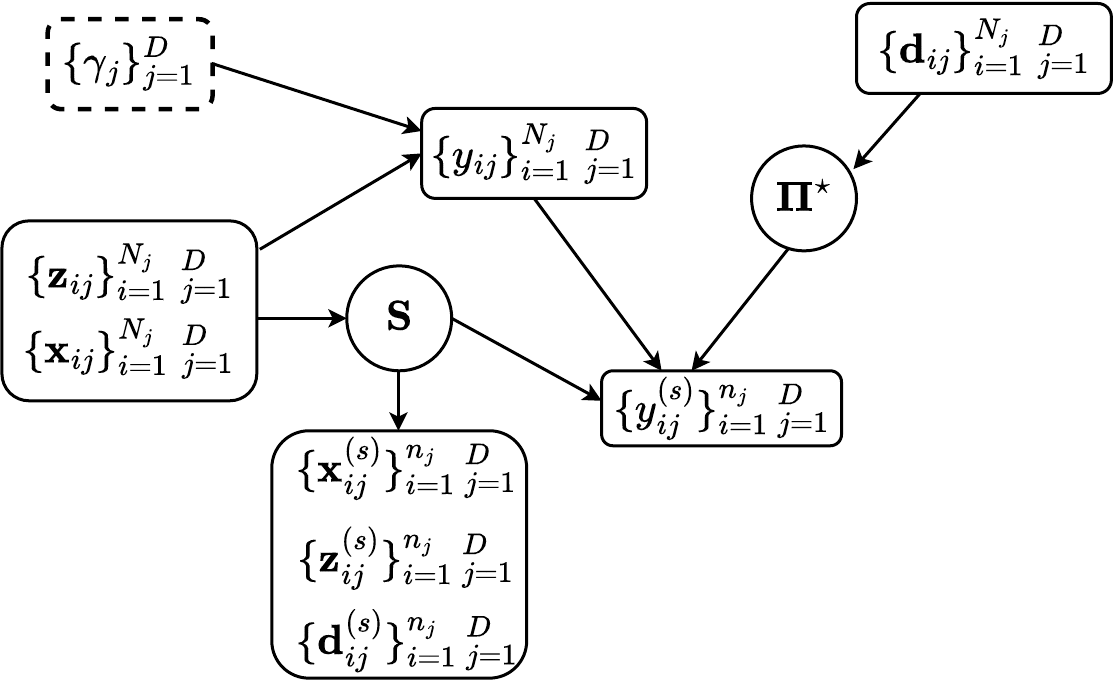}
\end{center}
\caption{Diagram summarizing the setting, its underlying quantities and their relationships. The random effects $\{ \bm{\gamma}_j \}_{j = 1}^D$ are in a dashed box since they arise only in the linear mixed effects SA model (cf.~$\S$\ref{saemodel_lmm})}\label{fig:setting}
\end{figure}
The nested error regression model is very popular in small area estimation since the original proposal of \cite{battese1988error}. In our notation, let us assume that the following model holds at the population level: 
\begin{align}\label{eq:unitlevel_SA}
\begin{split}
&y_{ij}|\M{x}_{ij}, \M{z}_{ij}, \bm{\gamma}_j \overset{\text{ind.}}{\sim} N(\M{x}_{ij}^{\T}
   \bm{\beta}_{*} + \M{z}_{ij}^{\T} \bm{\gamma}_j, \sigma_{*}^2),
   \quad i=1,\ldots,N_j, \quad j=1,\ldots,D,\\
   &\{ \bm{\gamma}_j \}_{j = 1}^D \overset{\text{i.i.d.}}{\sim}
   N_q(\bm{0}, \bm{\Sigma}_*),  
 \end{split}  
\end{align}
 for area-specific random effects $\{ \bm{\gamma}_j \}_{j = 1}^D$ and unknown parameters $\bm{\beta}_* \in \R^p$, $\sigma_*^2 > 0$,
   and $\bm{\Sigma}_* \in \R^{q \times q}$ positive definite.

Given the assumptions on sampling and linkage as stated in $\S$\ref{assumptions}, we have the following for every sampled unit: 
\begin{align}\label{eq:normal_mixture}
\begin{split}
    &y_{ij}^{(s)}|\M{x}_{ij}^{(s)},\M{z}_{ij}^{(s)}, \bm{\gamma}_j,
    \{ m_{ij}^{(s)}=0 \} \sim N(\M{x}_{ij}^{(s)\T}
   \bm{\beta}_{*} + \M{z}_{ij}^{(s)\T} \bm{\gamma}_j, \sigma_{*}^2) \\ 
   &y_{ij}^{(s)}|\M{x}_{ij}^{(s)}, \M{z}_{ij}^{(s)}, \bm{\gamma}_j, \{ m_{ij}^{(s)}=1 \} \sim g \\
   &\p(m_{ij}^{(s)} = 1 | \M{d}_{ij}^{(s)}) = h(\M{d}_{ij}^{(s)};\bm{\alpha}_*),        \qquad   1 \leq i \leq n_j, \; 1 \leq j \leq D. 
\end{split}
\end{align}
Note that the first two lines of display \eqref{eq:normal_mixture} represents a a two-component mixture model corresponding to the values of the latent mismatch indicators $\{ m_{ij}^{(s)} \}$ such that for correctly matched units, the standard linear mixed effects SA model \eqref{eq:unitlevel_SA} holds whereas for incorrectly matched units a ``null model" as specified by the density $g$ derived in \eqref{eq:marginal_ultimate} holds. 
\vskip1ex
\noindent {\bfseries Complete Data (Composite) Likelihood}. Let\footnote{To avoid notational clutter, we omit the superscript $^{(s)}$ in $\lambda_j$, $1 \leq j \leq D$} $\lambda_j = \{1 \leq i \leq n_j: m_{ij}^{(s)} = 0 \}_{j = 1}^D$ and $\{ \overline{\lambda}_j \}_{j = 1}^D = \{ \{1,\ldots,n_j\} \setminus \lambda_j \}_{j = 1}^D$. In accordance with terminology from the literature on missing data problems \citep{LittleRubin2019, Kim2021}, we consider the complete data likelihood associated with the mixture model \eqref{eq:normal_mixture} 
and the missing data $\{ \lambda_j \}_{j = 1}^D$ and $\{ \bm{\gamma}_j \}_{j = 1}^D$. We obtain (cf.~Appendix \ref{app:composite_likelihood_derivation} for a step-by-step derivation): 
\begin{align}
L(\bm{\beta}, \sigma^2, \bm{\Sigma}, \bm{\alpha}|\{ \lambda_j, \, \bm{\gamma}_j \}_{j = 1}^D) &= \prod_{j = 1}^D \bigg\{  (2 \pi \sigma^2)^{-|\lambda_j|/2} \exp \left(-\frac{\nnorm{\M{y}_{\lambda_j}^{(s)} - \M{X}_{\lambda_j}^{(s)} \bm{\beta} - \M{Z}_{\lambda_j}^{(s)} \bm{\gamma}_j}_2^2}{2 \sigma^2} \right) \times \notag \\
   & \quad \qquad \times \left[ |\bm{\Sigma}|^{-1/2} (2 \pi)^{q/2} \exp \left(-\frac{1}{2} \bm{\gamma}_j^{\T} \bm{\Sigma}^{-1} \bm{\gamma}_j \right) \right] \times \prod_{i \in \overline{\lambda}_j} g(y_{ij}^{(s)}) \times \notag \\
   &\quad \qquad \times \prod_{i \in \lambda_j} \underbrace{\p(m_{ij}^{(s)} = 0 | \M{d}_{ij}^{(s)}; \bm{\alpha})}_{1 - h(\M{d}_{ij}^{(s)};\bm{\alpha})}  \times \prod_{i \in \overline{\lambda}_j} \underbrace{\p(m_{ij}^{(s)} = 1 | \M{d}_{ij}^{(s)}; \bm{\alpha})}_{h(\M{d}_{ij}^{(s)};\bm{\alpha})} \bigg\}, \label{eq:pseudolikelihood}
\end{align}
where subscripts $\lambda_j$ indicate the associated sub-vectors and row submatrices, respectively. For now, $g$ is assumed to be fixed even though $g$ is generally unknown; This aspect revisited in $\S$\ref{subsec:EMalg} below\footnote{Since $g$ is assumed to be independent of linkage error (cf.~Eq.~\eqref{eq:non_informative}), its estimation does not require any specific adjustment.}. We note that 
\eqref{eq:pseudolikelihood} is more adequately referred to as a {\em composite likelihood} \citep{Lindsay1988, Varin2011} as opposed to a (proper) likelihood. This is because \eqref{eq:pseudolikelihood} may involve products over terms that are not independent. Specifically, the likelihood contributions involving the mismatch indicators are generally dependent since they are derived from a random permutation. As a simple example, consider a random permutation $\pi$ on $\{1,2\}$ that swaps the order of the 
two indices with probability $\alpha \in (0,1)$. In this case, $(1 - \alpha) = \p(\pi(1) = 1, \, \pi(2) = 2) \neq \p(\pi(1) = 1) \cdot \p(\pi(2) = 2) = (1-\alpha)^2$. This example relates to record linkage when one-to-one matchings are enforced: given two records $a$ and $b$ in the first file with correct matches $a'$ and $b'$ in the second file, respectively, then incorrectly matching $a$ to $b'$ will cause $b$ to be mismatched as well since the requirement of a one-to-match entails that (i) at least one match has to found for $b$ and (ii) $b$ cannot match to $b'$ since none of the records can appear
in more than one linked pair. 

Regardless, existing frameworks for composite likelihood inference justify the use of \eqref{eq:pseudolikelihood}: in a nutshell, under suitable regularity conditions estimators defined as maximizers of products of valid marginal likelihood contributions constructed from data subsets of any size (individual observations, pairs of observations, etc.) with possible dependencies across those subsets enjoy asymptotic properties similar to conventional maximum likelihood estimators; their asymptotic covariance matrix exhibits the usual ``sandwich form" that arises in the more general setting of $M$- and $Z$-estimators \citep[e.g.,][]{Stefanski2002}. 

The benefit of the composite likelihood \eqref{eq:pseudolikelihood} is that it yields a concise expression that retains most of the structure of the conventional linear mixed model likelihood and enables computationally tractable inference as detailed in $\S$\ref{subsec:EMalg} and Appendix \ref{app:EMalg}.  
\subsection{The linear M-quantile model}\label{saemodel_mq}
Similarly to the more popular quantile regression \citep{Koenker1978}, M-quantile (MQ) regression \citep{Breckling1988} is a semi-parametric method where M-estimation ideas are introduced in quantile regression to make it more flexible in terms of the balance to strike between robustness and efficiency. \cite{yu2001bayesian} use the asymmetric Laplace distribution to obtain a parametric representation of quantile regression. Along the same line, for the purposes of this paper we present the MQ regression model using a working likelihood introduced by \cite{Bianchi2018}. Let us assume that at the population level the following model holds:
\begin{equation}\label{w_gali_lik}
    y_{ij}|\M{x}_{ij} \sim \text{GALI}_{\tau_j}(\M{x}_{ij}^{\T} \bm{\beta}_{*\tau_j}, \sigma_{*\tau_j}),
\quad 1 \leq i \leq N, \; 1  \leq j \leq D,
\end{equation}
for some sequence $(\tau_j)_{j = 1}^D \subset (0,1)$, where the $\text{GALI}_{\tau}(\mu, \sigma)$ distribution with $\tau \in (0,1), \, \mu \in \R \: \text{and} \, \sigma > 0$ is characterized by the density function
\begin{equation} \label{GALI_density}
f^G_\tau(y; \mu, \sigma)= \frac{1}{\sigma B_\tau(\mu,\sigma)} \exp{\left\{ -\rho_\tau\left(\frac{y-\mu}{\sigma} \right) \right\}}, ~~~-\infty <y <+\infty.
\end{equation}
In Eq.~\eqref{GALI_density}, $B_{\tau} = \int_{-\infty}^{+\infty} \frac{1}{\sigma} \exp\left\{ -\rho_\tau\left(\frac{y-\mu}{\sigma} \right) \right\} dy < +\infty$, and $\mu$ and $\sigma$ are location and scale parameters, respectively. The function $\rho_\tau: \R \rightarrow \R+$ inside the exponential is a tilted loss function $u \mapsto \rho_{\tau}(u) =|\tau-I(u<0)|\rho(u)$ for a non-negative function $\rho$ that is twice differentiable with $\rho(0) = 0$.

In line with the notation and assumptions introduced in $\S$\ref{assumptions} and the basic idea of the approach formulated in $\S$\ref{saemodel_lmm}, at the sample level we postulate that:
\begin{align}\label{eq:GALI_mixture}
\begin{split}
&y_{ij}^{(s)}|\M{x}_{ij}^{(s)},\{m_{ij}^{(s)}=0\}\sim \text{GALI}_{\tau_j}(\M{x}_{ij}^{(s) \,
    \T} \bm{\beta}_{*\tau_j}, \sigma_{*\tau_j}) \\
&y_{ij}^{(s)}|\M{x}_{ij}^{(s)},\{m_{ij}^{(s)}=1\}\sim \text{GALI}_{\tau_j}(\theta_{*\tau_j}, \varsigma_{*\tau_j}) \\
&\p(m_{ij}^{(s)} = 1 | \M{d}_{ij}^{(s)}) = h(\M{d}_{ij}^{(s)};\bm{\alpha}_*),  \qquad   1 \leq i \leq n_j, \; 1 \leq j \leq D.   
\end{split}
\end{align}
In the second line of \eqref{eq:GALI_mixture}, $\theta_{*\tau_j}$ denotes the $\tau_j$-M-quantile of the marginal distribution of the sampled $y$'s and $\varsigma_{*\tau_j}$ denotes the associated scale parameter. 
This constitutes the second departure from the mixture model specified in \eqref{eq:normal_mixture} that replaces the marginal density of the sampled $y$'s $g$ by a simpler parametric (GALI) model that -- from a regression viewpoint -- can be interpreted as an intercept-only MQ regression model. This simplification is motivated by the fact that already the GALI model for correct matches is used as a working likelihood (as opposed to a likelihood deduced from a generative model such as \eqref{eq:unitlevel_SA}),    
hence an elaborate nonparametric specification for incorrect matches is considered disproportionate from the perspective of inference. Further note that different from \eqref{eq:normal_mixture}, the first line in \eqref{eq:GALI_mixture} does not involve random effects. Instead, the area-specific variation not accounted for by covariates enters by means of the area-specific quantiles $(\tau_j)_{j = 1}^D$ that are determined in a data-driven manner as explained in $\S$\ref{MQ_predicor_predictor} below.     

With similar considerations as in $\S$\ref{saemodel_lmm}, the complete data composite (working) likelihood for the linear MQ regression model for a given $\tau$ can be expressed as
\begin{eqnarray}
L(\bm{\beta}_{\tau}, \sigma_{\tau}, \bm{\alpha}|\{ \lambda_j\}_{j = 1}^D) 
&=& \prod_{j=1}^{D}\Bigg [\prod_{i \in \lambda_j}  \underbrace{\p(m_{ij}^{(s)} = 0 | \M{d}_{ij}^{(s)};\bm{\alpha})}_{1 - h(\M{d}_{ij}^{(s)};\bm{\alpha})} \cdot \frac{1}{\sigma_{\tau } B_\tau}  \exp{\left\{ -\rho_\tau\left(\frac{y_{ij}^{(s)}-\M{x}_{ij}^{(s) \,
    \T} \bm{\beta}_{\tau}}{\sigma_{\tau}} \right) \right\}} \notag \\
    &\times& \prod_{i \in \overline{\lambda}_j} \underbrace{\p(m_{ij}^{(s)} = 1 | \M{d}_{ij}^{(s)};\bm{\alpha})}_{h(\M{d}_{ij}^{(s)};\bm{\alpha})}
    \cdot \frac{1}{\varsigma_{\tau } B_\tau }\exp{\left\{ -\rho_\tau\left(\frac{y_{ij}^{(s)}-
    \theta_{\tau}}{\varsigma_{\tau}} \right) \right\}}\Bigg ]. 
\end{eqnarray}
Given the assumptions in $\S$\ref{assumptions}, in particular non-informativeness of the linkage error, the parameters $(\theta_{\tau}, \varsigma_{\tau})$ associated with the second component of the mixture can be (for the moment) treated as fixed, although in practice they will not and are thus to be replaced by MQ-based location and scale estimators.

\section{The LMM-based predictor}\label{LMM_predictor}
In the next two sections, we study the approaches outlined in $\S$\ref{saemodel_lmm} and $\S$\ref{saemodel_mq} and the resulting predictors for the area means in more detail. We start with the former approach. Before discussing the predictor and estimation of its MSE, we first sketch our computational algorithm for estimating the underlying parameters, with detailed computations relegated to Appendix \ref{app:EMalg}.    
\subsection{Computational algorithm}\label{subsec:EMalg}
Estimation of the parameters $\bm{\theta}_* = (\bm{\beta}_*, \sigma_*^2,  \bm{\Sigma}_*, \bm{\alpha}_*)$ is based on 
the {\em integrated} complete data (composite)-likelihood arising from \eqref{eq:pseudolikelihood} via integration 
over the latent variables $\{ (\lambda_j, \bm{\gamma}_j) \}_{j = 1}^D$, i.e., 
\begin{align}
&L(\bm{\beta}, \sigma^2, \bm{\Sigma}, \bm{\alpha}) = \int \hspace*{-.7ex}\sum_{\{ \ell_1 \} \in \mc{P}_1} \ldots \sum_{\{ \ell_D \} \in \mc{P}_D}  \hspace*{-.5ex}\Big\{ L(\bm{\beta}, \sigma^2, \bm{\Sigma}, \bm{\alpha} | \{ \ell_j, \bm{\gamma}_j \}_{j = 1}^D) \; \times \prod_{j = 1}^D \{ \p(\lambda_j = \ell_j) \; f(\bm{\gamma}_j) \} \Big\}  \notag \\
&\qquad \qquad \qquad \qquad \qquad \qquad \qquad \qquad \;\; d{\bm{\gamma}}_1 \ldots d\bm{\gamma}_D. \label{eq:integrated_likelihood}
\end{align}
where the summations are over all possible realizations $\{ \ell_j \}$ of the configurations of mismatches in area $j$, which is given by $\mc{P}_j$ -- the power set of  $\{1,\ldots,n_j \}$, $1 \leq j \leq D$. The representation \eqref{eq:integrated_likelihood} indicates that direct maximization of $L$ poses a significant computational challenge; in fact, the mere \emph{evaluation} of the integrated likelihood alone is intricate. For this reason, we resort to the EM algorithm \citep{Dempster1977}. The E-step requires considerable computational effort given that the latent $\{ \lambda_j \}$ take values in the power sets $\{ \mc{P}_j \}_{j = 1}^D$. In order to ensure scalability of the E-step to larger sample sizes $\{ n_j \}$, we develop a Monte Carlo approximation based on separable updates of the individual
mismatch indicators $\{ m_{ij}^{(s)} \}$ via Gibbs sampling \citep[e.g.][]{Robert2013}. Note that one-by-one updating of the $\{ m_{ij}^{(s)} \}$ in the exact E-step is not possible given the dependence of the $\{ y_{ij} \}_{i = 1}^{n_j}$ induced by $\bm{\gamma}_j$, $1 \leq j \leq D$, and the fact that the E-step involves the joint expectation of $\{ (\lambda_j, \bm{\gamma}_j) \}$ given 
the data and current values of the parameters. 

By comparison, the resulting M-step is straightforward. The update for $\bm\beta$ turns out to reduce to a weighted least
squares problem. The updates for the variance components $\bm{\Sigma}$ and $\sigma^2$ have simple and computationally tractable closed form solutions. Finally, the update for $\bm{\alpha}$ may require iterative algorithms depending on the complexity of the function $h$; in some cases, closed form updates are obtained. Detailed derivations of the E-step (both in the exact case and with the use of the proposed Monte-Carlo approximation) and the M-step can be found in Appendix \ref{app:EMalg}.
\vskip1ex
{\em Further implementation details}. So far, we have assumed that the density $g$ (cf.~Eq.~\eqref{eq:marginal_ultimate}) is known which is not the case in practice. Note that estimation of $g$ -- the
marginal PDF of the sampled $y$'s -- is not affected by mismatch error and thus does not require the use of specific strategies to deal with such error. Straightforward approaches include kernel density estimation (KDE) or the use of 
(simplified) parametric models, e.g., $\{ y_{ij}^{(s)} \} \sim N(\mu_g, \sigma_g^2)$ if the total number of sampled 
units is too small for the use of KDE. According to the expression for $g$ given the linear mixed model \eqref{eq:unitlevel_SA} provided at the end of Appendix \ref{app:composite_likelihood_derivation} has a finite Gaussian mixture representation. While the locations and scales of this mixture \eqref{eq:marginal_ultimate} are unknown, they
can be substituted given pilot estimators for $\bm{\beta}_*$, $\bm{\Sigma}_*$, and $\sigma_*^2$. Possible pilot estimators
include the naive estimator (ignoring mismatch error) given the linear mixed model \eqref{eq:unitlevel_SA} or one of the proposals developed in \cite{salvati2021} if estimates for block-wise mismatch rates are available. The resulting 
estimate of $g$, say $\wh{g}^{(0)}$, can be used in a first round of EM iterations as detailed, yielding estimators 
$(\wh{\bm{\beta}}, \wh{\bm{\Sigma}}, \wh{\sigma}^2)$ than can in turn be used to update the locations and scales in 
the mixture \eqref{eq:marginal_ultimate}, yielding a refined estimate $\wh{g}^{(1)}$ to be used in another round of 
EM iterations to obtain refined estimates of the parameters. This process can be continued iteratively.

Empirical results on the performance of the proposed approach reported in this paper do not make use of the described
iterative scheme since vanilla KDE with a Gaussian kernel and default choices of the bandwidth achieved satisfactory
performance. Similarly, initialization of the EM iterations with the naive (REML or ML) estimates in line with \eqref{eq:unitlevel_SA} worked well even though improved initialization could be attempted. 

\subsection{The predictor}
Given the linear mixed model \eqref{eq:unitlevel_SA}, the usual estimator of the SA means is given by 
\begin{equation}\label{eq:SAmeans_estimator}
\wh{\overline{y}}_{j} = \frac{1}{N_j} \left( \sum_{i = 1}^{n_j} y_{ij}^{(s)} + \sum_{k \notin \{ s_{ij}  \}}  \left( \M{x}_{kj}^{\T} \wh{\bm\beta} + \M{z}_{kj}^{\T} \wh{\bm{\gamma}}_j \right) \right), \quad 1 \leq j \leq D,  
\end{equation}
where $\wh{\bm{\beta}}$ is an estimator of $\bm{\beta}_*$ and $\{ \wh{\bm\gamma}_j \}_{j = 1}^D$ are predictors 
of $\{ \bm \gamma_j \}_{j = 1}^D$, and we recall that $\{ s_{ij} \}_{i = 1}^{n_j}$ represent the indices of the 
sampled unit in area $j$, $1 \leq j \leq D$. In the absence of linkage error, the above estimator is the best unbiased linear predictor (BLUP) of the SA means $\{ \overline{y}_{j} \}_{j=1}^D$ under model \eqref{eq:unitlevel_SA} if 
$\bm{\Sigma}_*$ and $\sigma_*^2$ are known and the associated best linear unbiased estimator (BLUE) for $\bm{\beta}_*$ and BLUPs for $\{ \bm{\gamma}_j \}_{j = 1}^D$ are employed \citep[cf.,][$\S$7.1.3]{rao2015}. 

Note that the first summand in \eqref{eq:SAmeans_estimator} can be expressed as $(n_j/N_j) \overline{y}_{j}^{(s)}$, where
$\overline{y}_{j}^{(s)}$ denotes the mean of the samples $\{ y_{ij}^{(s)} \}$ in area $j$, $1 \leq j \leq D$. The contribution of these terms are minor whenever the corresponding sampling ratios $\{ n_j / N_j \}$ are small. In the presence of linkage error (possibly across areas) affecting the $\{ y_{ij}^{(s)} \}$, we drop the first term altogether, and instead use 
\begin{equation}\label{eq:predictor_lmm}
\wh{\overline{y}}_{j} = \frac{1}{N_j} \sum_{i = 1}^{N_j} \big( \M{x}_{ij}^{\T} \wh{\bm{\beta}} + \M{z}_{ij}^{\T} \wh{\bm{\gamma}}_j \big) = \overline{\M{x}}_j^{\T} \wh{\bm{\beta}} + \overline{\M{z}}_j^{\T} \wh{\bm{\gamma}}_j, \quad 1 \leq j \leq D,
\end{equation}
where $\overline{\M{x}}_j \coloneq \frac{1}{N_j} \sum_{i = 1}^{N_j} \M{x}_{ij}$, $\overline{\M{z}}_j \coloneq \frac{1}{N_j} \sum_{i = 1}^{N_j} \M{z}_{ij}$, 
 $\wh{\bm{\beta}}$ is the argument in $\bm{\beta}$ that maximizes the integrated complete data (composite) likelihood \eqref{eq:integrated_likelihood}, and 
\begin{align}\label{eq:gamma_hat}
\begin{split}
\wh{\bm\gamma}_j = \E[\bm{\gamma}_j | \mc{D}; \wh{\bm{\theta}}] &= \sum_{\ell_j \in \mc{P}_j} \p(\lambda_j = \ell_j| \mc{D}; \wh{\bm{\theta}}) \bm\mu_{\ell_j}(\wh{\bm{\theta}}), \quad \bm{\mu}_{\ell_j}(\wh{\bm{\theta}}) \coloneq \E[\bm{\gamma}_j | \mc{D}, \{ \lambda_j = \ell_j \}; \wh{\bm{\theta}}], \\
&= \sum_{\ell_j \in \mc{P}_j} \p(\lambda_j = \ell_j| \mc{D}; \wh{\bm{\theta}}) \big\{ \underbrace{\wh{\bm{\Sigma}} \M{Z}_{\ell_j}^{(s)\T} \left( \M{Z}_{\ell_j}^{(s)} \wh{\bm{\Sigma}} \M{Z}_{\ell_j}^{(s)\T} + \wh{\sigma}^{2}  I_{|\ell_j|}\right)^{-1} (\M{y}_{\ell_j}^{(s)} - \M{X}_{\ell_j}^{(s)} \wh{\bm{\beta}})}_{\bm{\mu}_{\ell_j}(\wh{\bm{\theta}})} \big\},
\end{split}
\end{align}
$1 \leq j \leq D$. Here, $\p(\ldots | \mc{D}; \bm{\theta}_0)$ and $\E[\ldots | \mc{D}; \bm{\theta}_0]$ denote probabilities and expectations taken conditional on the observed data $\mc{D}$ and assuming that the model parameters are equal to some $\bm{\theta}_0 = (\bm{\beta}_0, \sigma_0^2,  \bm{\Sigma}_0, \bm{\alpha}_0))$ contained in the underlying parameter space. The terms arising in \eqref{eq:gamma_hat} are already calculated in the E-step of the final EM iteration, in both the exact as well as in the Monte-Carlo variant of the EM scheme detailed in Appendix \ref{app:EMalg}, which in particular provides a specific expression for the probabilities inside the sum.

Observe that if $\bm{\theta}_*$ were known, then substitution of $\wh{\bm{\theta}}$ by $\bm{\theta}_*$ in \eqref{eq:gamma_hat} yields an (MSE-) optimal unbiased estimators of the $\{ \bm{\gamma}_j \}_{j = 1}^D$ that, however, is non-linear in the $\{ y_{ij}^{(s)} \}$ as can be verified from the specific form of the $\{ \p(\lambda_j = \ell_j| \mc{D}; \wh{\bm{\theta}}): \; \ell_j \in \mc{P}_j, \; 1 \leq j \leq D\}$. With $\wh{\bm{\theta}}$ being a consistent estimator of $\bm{\theta}_*$ (cf.~\eqref{eq:asymptotic_composite} in $\S$\ref{sec:MSEEBLUP} below), asymptotic unbiasedness of 
$\{ \wh{\bm{\gamma}}_j \}_{j = 1}^D$ as $D \rightarrow \infty$ follows from the continuous mapping theorem. 

\subsection{MSE estimation}\label{sec:MSEEBLUP}

{\bfseries MSE estimation with known parameters}. 
In order to assess the MSE of the predictor in the previous subsection, let us first consider the situation in which the parameters $\bm{\theta}_*$ are fully known, corresponding to the term $g_1$ in the usual MSE decomposition $g_1 + g_2 + g_3$ in the literature \citep[cf.,][$\S$6.2]{rao2015}. In the computation of $g_1$ the only source of variability in the predictor \eqref{eq:predictor_lmm} arises from
the random effects $\{ \bm{\gamma}_j \}_{j = 1}^D$, and thus 
\begin{equation}\label{eq:MSE_allknown}
\text{MSE}(\wh{\overline{y}}_j) = \overline{\M{z}}_j^{\T} \cov(\bm{\gamma}_j | \mc{D}; \bm{\theta}_*) \overline{\M{z}}_j. 
\end{equation}
The covariance in the previous display an be further decomposed as 
\begin{align}
\cov(\bm{\gamma}_j | \mc{D}; \bm{\theta}_*) &= \E_{\lambda_j}[\cov(\bm{\gamma}_j | \lambda_j, \mc{D}; \bm{\theta}_*)] + \cov_{\lambda_j}(\E[\bm{\gamma}_j | \lambda_j, \mc{D}; \bm{\theta}_*]) \label{eq:gamma_cov}\\
&= \sum_{\ell_j \in \mc{P}_j} \p(\lambda_j = \ell_j | \mc{D}; \bm{\theta}_*) \left\{ \bm{\Sigma}_{\ell_j}  (\bm{\theta}_*) + \left( \bm{\mu}_{\ell_j}(\bm{\theta}_*) - \wh{\bm\gamma}_j(\bm{\theta}_*) \right)\left( \bm{\mu}_{\ell_j}(\bm{\theta}_*) - \wh{\bm\gamma}_j(\bm{\theta}_*) \right)^{\T} \right\} \notag,
\end{align}
where $\wh{\bm{\gamma}}_j(\bm{\theta}_*)$ and $\{ \bm{\mu}_{\ell_j}(\bm{\theta}_*) \}_{\ell_j \in \mc{P}_j}$ are defined analogously to $\wh{\bm{\gamma}}_j = \wh{\bm{\gamma}}_j(\wh{\bm\theta})$ 
and $\{ \bm{\mu}_{\ell_j}(\wh{\bm{\theta}}) \}_{\ell_j \in \mc{P}_j}$ in \eqref{eq:gamma_hat}, respectively, and
\begin{equation*}
\bm{\Sigma}_{\ell_j}(\bm{\theta}_*) = \cov(\bm{\gamma}_j | \mc{D}, \{ \lambda_j = \ell_j \}; \bm{\theta}_*) = \bm{\Sigma}_* - \bm{\Sigma}_* \M{Z}_{\ell_j}^{\T} \left( \M{Z}_{\ell_j} \bm{\Sigma}_* \M{Z}_{\ell_j}^{\T} + \sigma_*^{2} I_{|\ell_j|}\right)^{-1} \M{Z}_{\ell_j} \bm{\Sigma}_*, \quad \ell_j \in \mc{P}_j.
\end{equation*}
The expression for the MSE \eqref{eq:MSE_allknown} can be evaluated accordingly, either exactly or approximately along the lines of the Monte-Carlo EM approach in Appendix \ref{subsec:MCEM}. 
\vskip1ex
\noindent {\bfseries MSE estimation with known variance components}. Next, we suppose that all parameters except for the variance components $\bm{\Sigma}_*$ and $\sigma_*^2$ are unknown. We aim to approximate the MSE attributable to the random effects and the estimation of 
$\bm{\beta}_*$ and $\bm{\alpha}_*$, corresponding to the term $g_1 + g_2$ in the aforementioned MSE decomposition. 
For this purpose, we infer the large-sample distribution of $(\wt{\bm{\beta}}, \wt{\bm{\alpha}})$ denoting the estimator of $(\bm{\beta}_*, \bm{\alpha}_*)$ given $\bm{\Sigma}_*$ and $\sigma_*^2$ based on \eqref{eq:integrated_likelihood}. We then generate samples from that distribution to estimate the MSE akin to Multiple Imputation \citep{LittleRubin2019}. Specifically, letting 
$\{ (\wt{\bm{\beta}}^{[m]}, \wt{\bm\alpha}^{[m]}) \}_{m = 1}^M$ denote the thus generated samples, the MSE is estimated as 
\begin{align}
\begin{split}
\wh{\text{MSE}}(\wh{\overline{y}}_j) &= \E_{M}[ \overline{\M{z}}_j^{\T} \cov(\bm{\gamma}_j | \mc{D}; \wt{\bm{\beta}}^{[m]}, \wt{\bm\alpha}^{[m]}, \bm\Sigma_*, \bm{\sigma}_*^2) \overline{\M{z}}_j] + \\ \quad &+\var_M\big(\{ \overline{\M{x}}_{j}^{\T} \wt{\bm{\beta}}^{[m]}  + \overline{\M{z}}_j^{\T} \E[\bm{\gamma}_j | \mc{D}; \wt{\bm{\beta}}^{[m]}, \wt{\bm\alpha}^{[m]}, \bm\Sigma_*, \bm{\sigma}_*^2]\}_{m = 1}^M \big) \label{eq:MSE_MC}\\
&= \text{Within-Sample-Variation} + \text{Between-Sample-Variation} 
\end{split}
\end{align}
$1 \leq j \leq D$, where $\E_{M}$ and $\var_M$ refer to the empirical mean and variance, respectively, with respect to the sample of size $M$, and $\cov(\bm{\gamma}_j | \ldots)$ and $\E[\bm{\gamma}_j | \ldots]$ are defined analogously to \eqref{eq:gamma_hat}
and \eqref{eq:gamma_cov}, respectively. 

In order to find the large-sample distribution of $(\wt{\bm{\beta}}, \wt{\bm{\alpha}})$, we leverage properties of composite maximum likelihood estimators \cite[cf.][]{Varin2011} in a super-population setting. Specifically, it holds that 
\begin{equation}\label{eq:asymptotic_composite}
\sqrt{n} \left( \begin{array}{c}
                \wt{\bm{\beta}} - \bm{\beta}_* \\
                \wt{\bm{\alpha}} - \bm{\alpha}_*
                \end{array} \right) \overset{D}{\rightarrow} N(\M{0}, \, \E[\nabla^2 \mc{L}(\bm{\theta}_*)]^{-1}\, \cov(\mc{L} (\bm{\theta}_*))  \,  \E[\nabla^2 \mc{L}(\bm{\theta}_*)]^{-1}   \Bigg),
\end{equation}
where $\overset{D}{\rightarrow}$ denotes convergence in distribution, $\mc{L}(\cdot)$ denotes the negative logarithm of of the integrated complete data (composite)-likelihood \eqref{eq:integrated_likelihood}. Moreover, $\nabla$ and $\nabla^2$ here denote the gradient and Hessian with respect to $(\bm{\beta}, \bm{\alpha})$. Since the direct evaluation of gradients and Hessians of $\ell$ is unwieldy and since $(\wt{\bm{\beta}}, \wt{\bm{\alpha}})$ are computed via the EM algorithm anyway, we resort to useful identities in \cite{Louis1982} concerning the score and observed information associated with incomplete data likelihood functions to estimate the covariance matrix. To be specific, we use 
\begin{align}\label{eq:covariance_louis}
\begin{split}
\widehat{\E[\nabla^2 \mc{L}(\bm{\theta}_*)]}  &= \frac{1}{n} \sum_{j = 1}^D \nabla^2 \mc{L}_j(\wt{\bm{\theta}}) = \frac{1}{n} \sum_{j = 1}^D \{ \E_{\lambda_j, \bm{\gamma}_j}[\nabla^2 \mc{L}_j^{\textsf{c}}(\wt{\bm{\theta}}) | \mc{D}; \wt{\bm{\theta}}] - \cov_{\lambda_j, \bm{\gamma}_j}(\nabla \mc{L}_j^{\textsf{c}}(\wt{\bm{\theta}})|\mc{D}; \wt{\bm{\theta}}) \} \\
\widehat{\cov(\nabla \mc{L} (\bm{\theta}_*))} &= \frac{1}{n} \sum_{j = 1}^D \nabla \mc{L}_j(\wt{\bm{\theta}}) \nabla \mc{L}_j(\wt{\bm{\theta}})^{\T} = \frac{1}{n} \sum_{j = 1}^D \E_{\lambda_j, \bm{\gamma}_j}[\nabla \mc{L}_j^{\textsf{c}}(\wt{\bm{\theta}})|\mc{D}; \wt{\bm{\theta}}] \, \E_{\lambda_j, \bm{\gamma}_j}[\nabla \mc{L}_j^{\textsf{c}}(\wt{\bm{\theta}})|\mc{D}; \wt{\bm{\theta}}]^{\T}, 
\end{split}
\end{align}
where here and in the sequel, we use $\wt{\bm{\theta}} = (\wt{\bm{\beta}}, \wt{\bm{\alpha}}, \bm{\Sigma}_*, \sigma_*^2)$, and the second identities in both equations in \eqref{eq:covariance_louis} are in virtue of the results in \cite{Louis1982}, with $\mc{L}_j^{\textsf{c}}$ denoting the $j$-th term in the complete data negative log-likelihood given in \eqref{eq:completedata_negloglik}, $1 \leq j \leq D$. Evaluating the expectations and covariances with respect to the latent variables $\{\lambda_j, \bm{\gamma}_j \}_{j = 1}^D$ in the right hand sides of \eqref{eq:covariance_louis} in closed form is possible in principle, but inevitably requires tedious calculations involving a large number of terms\footnote{It can be verified that reproducing known expressions for the observed information for the {\em standard} linear mixed model (i.e., without mismatch error) based on Louis' identity requires significant effort.}; apart from that, as noted above, such closed form expressions quickly become intractable with growing sample sizes $\{ n_j \}_{j = 1}^D$. Therefore, we propose approximate evaluation of these expectations and covariances via Monte Carlo integration following the sampling scheme in $\S$\ref{subsec:MCEM}. Further details such as expressions for $\{ \nabla \mc{L}_j^{\textsf{c}} \}$ are $\{ \nabla^2 \mc{L}_j^{\textsf{c}} \}$ are provided in Appendix \ref{app:MSE_eblup}.
\vskip1ex
\noindent {\bfseries MSE estimation in the general case}. Since the variance components are typically unknown, we replace them with the corresponding estimates produced by our approach. We then apply the scheme outlined in the previous paragraph, ignoring the uncertainty associated with the variance components. We show empirically that the resulting MSE estimates are still accurate. Given the inherently non-linear nature of the proposed predictor, MSE estimation is associated with significant technical challenges, and accounting for uncertainty from estimation of the variance components is considered out of the scope of the present paper and left for future work.

\section{The MQ-based predictor}\label{MQ_predictor}

\subsection{The predictor}\label{MQ_predicor_predictor}
In the application of MQ regression to small area estimation, the construction of the predictor hinges on the unit-level MQ coefficients and their area-level averages that are used to capture the unobserved heterogeneity at the area level. As noted in \cite{Chambers2006}, at the population level we can define observation-specific quantiles $\tau_{ij} \in (0,1)$ so that $y_{ij}=\M{x}_{ij}^{\T} \bm{\beta}_{\tau_{ij}}$. As MQ regression is based on a smooth loss function for each $y_{ij}$ we have exactly one such $\tau_{ij}$. The area-level population mean can be represented as
\begin{equation*}
\overline{y}_{j} = N_j^{-1} \left\{\sum_{ i = 1}^{n_j} y_{ij}^{(s)} +
\sum_{k \notin \{s_{ij}\}}\M{x}_{kj}^{\T} \bm{\beta}_{\tau_{kj}} \right\}.
\end{equation*}
In prediction, the unknown $\{ \tau_{ij} \}$ of the non-sampled units are replaced, in applications without linkage problems, by area-specific (average) quantiles $\tau_{j}=n_{j}^{-1} \sum_{k \in \{ s_{ij} \}} \tau_{kj}$ assuming that the individual $\{ \tau_{ij} \}$ are more alike within areas than they are between areas. 
In our context, once estimates of the model parameters $\{ \bm{\theta}_{*\tau} = \left(\bm{\beta}_{*\tau},\sigma_{*\tau}, \bm{\alpha}_{*\tau} \right): \, \tau \in T \subset (0,1)  \}$ are available, $\tau_{j}$ is estimated by $\widehat{\tau}_{j}=(\sum_{i = 1}^{n_j} \omega_{ij})^{-1}(\sum_{i = 1}^{n_j} \omega_{ij} \widehat{\tau}_{ij})$, with $\widehat{\tau}_{ij}$ defined by $y_{ij}^{(s)} =\M{x}_{ij}^{(s)\T}\widehat{\bm{\beta}}_{\widehat{\tau}_{ij}}$ and $\omega_{ij} =\E[1 - m_{ij}|\mc{D};\widehat{\bm{\theta}}_{\widehat{\tau}_{ij}}]$ as defined in \eqref{eq:EMweights_MQ} below. The purpose of the $\{ \omega_{ij} \}$ is to assign higher weights to sampled units that are deemed more likely to be correctly linked.

The proposed MQ-predictor for $\overline{y}_{j}$ is given by
\begin{equation} \label{mq_predictor}
\widehat{\overline{y}}_{j}^{\text{MQ}} = \frac{1}{N_j} 
\sum_{ i = 1}^{N_j} \M{x}_{ij}^{\T} \wh{\bm{\beta}}_{\widehat{\tau}_{j}}
= \overline{\M{x}}_j^{\T}\wh{\bm{\beta}}_{\widehat{\tau}_{j}}
,\quad 1 \leq j \leq D, 
\end{equation}
where parallel to \eqref{eq:predictor_lmm}, the term $\sum_{ i = 1}^{n_j} y_{ij}^{(s)}$ is dropped to account for the possibility that linkage error may occur across areas. The rightmost term in \eqref{mq_predictor} highlights that the computation of $\widehat{\overline{y}}_{j}^{\text{MQ}}$ only requires the knowledge of the population-level average $\overline{\M{x}}_j$ and rather than the individual $\{ \M{x}_{ij} \}$  for the whole population.

\subsection{Computational algorithm}
We first note that given the indicator variables $\{m_{ij}^{(s)}\}$ and $\tau \in (0,1)$, the complete data negative (composite) log-likelihood in $\bm{\theta}_{\tau} \coloneq \left(\bm{\beta}_{\tau},\sigma_{\tau}, \bm{\alpha}_{\tau} \right)$ under the $\text{GALI}_{\tau}$ model is given by 
\begin{eqnarray}\label{eq:loglike:mq}
\nonumber \sum_{j=1}^{D}\sum_{i=1}^{n_j} & &\left\{ (1-m_{ij}) \left(-\log{(1-h(\M{d}_{ij};\bm{\alpha}_{\tau}))}+\log{(\sigma_{\tau} B_\tau)} +\rho_\tau\left(\frac{y_{ij}-\M{x}_{ij}^{\T} \bm{\beta}_{\tau}}{\sigma_{\tau}} \right) \right) \right.\\
& &  + \, m_{ij} \left(-\log{(h(\M{d}_{ij};\bm{\alpha}_{\tau}))}+\log{(\varsigma_{\tau} B_\tau)} +\rho_\tau\left(\frac{y_{ij}-\theta_\tau}{\varsigma_{\tau}} \right) \right) \Bigg\}.
\end{eqnarray} 
where we have dropped the superscript $^{(s)}$ (as will we do for the rest of this section) to avoid confusion with an iteration counter. We recall from the discussion in $\S$\ref{saemodel_mq} that $\theta_{\tau}$ and $\varsigma_{\tau}$ are considered fixed. As for the first predictor, we adopt  
an EM algorithm for optimization, which however, is less involved. Given current iterates $\wh{\bm\theta}_{\tau}^{(t)} \coloneq (\widehat{\bm{\beta}}_{\tau}^{(t)}, \, \widehat{\sigma}_{\tau}^{(t)}, \, \widehat{\bm{\alpha}}_{\tau}^{(t)})$ at iteration $t$, the E-step is given by (cf.~also \cite{salvati2021}, $\S$5)
{\small \begin{align}\label{eq:EMweights_MQ}
    \omega_{ij}^{(t)} &:=  \E[1 - m_{ij} | \mc{D}; \wh{\bm\theta}_{\tau}^{(t)}] \notag\\
    &= \frac{\displaystyle\frac{(1-h(\M{d}_{ij};\widehat{\bm{\alpha}}_{\tau}^{(t)}))}{\wh{\sigma}_{\tau}^{(t)} B_\tau} \exp{\left\{ -\rho_\tau\left(\frac{y_{ij} -\M{x}_{ij}^{\T} \widehat{\bm{\beta}}_{\tau}^{(t)}}{\wh{\sigma}_{\tau}^{(t)}} \right) \right\}}}{\displaystyle\frac{(1-h(\M{d}_{ij};\widehat{\bm{\alpha}}_{\tau}^{(t)}))}{\wh{\sigma}_{\tau}^{(t)} B_\tau} \exp{\left\{ -\rho_\tau\left(\frac{y_{ij} -\M{x}_{ij}^{\T} \widehat{\bm{\beta}}_{\tau}^{(t)}}{\wh{\sigma}_{\tau}^{(t)}} \right) \right\}}+ \frac{h(\M{d}_{ij};\widehat{\bm{\alpha}}_{\tau}^{(t)})}{\varsigma_\tau B_\tau} \exp \left(-\rho_\tau\left(\frac{y_{ij}-\theta_\tau}{\varsigma_{\tau}} \right) \right)},
\end{align}}
where $\M{E}[\ldots|\mc{D};\wh{\bm\theta}_{\tau}^{(t)} ]$ denotes the expectation conditional on the observed data $\mc{D}$ assuming that the unknown parameters of the underlying distribution are given by $\wh{\bm\theta}_{\tau}^{(t)}$. Accordingly, in light of \eqref{eq:loglike:mq} the resulting M-step yields the following updates: 
\begin{eqnarray*}
   \widehat{\bm{\alpha}}_{\tau}^{(t+1)}& = &\argmin_{\bm{\alpha}_{\tau}}\left\{-\sum_{j=1}^{D}\sum_{i=1}^{n_j}\left[ \omega_{ij}^{(t)} \log{ \left( 1-h(\M{d}_{ij};\bm{\alpha}_{\tau})\right)} + (1-\omega_{ij}^{(t)}) \log \left( h(\M{d}_{ij};\bm{\alpha}_{\tau})\right)\right]\right\},\\
   \widehat{\bm{\beta}}_{\tau}^{(t+1)} &=&\argmin_{\bm{\beta}_{\tau}} \left\{ -\sum_{j=1}^{D}\sum_{i=1}^{n_j} \omega_{ij}^{(t)} \rho_\tau\left(\frac{y_{ij} -\M{x}_{ij}^{
    \T} \bm{\beta}_{\tau}}{\sigma^{(t)}_{\tau}} \right)\right\},\\
    \widehat{\sigma}_{\tau}^{(t+1)} &=&\argmin_{\sigma_{\tau}} \left\{-\log (\sigma_{\tau}) \sum_{j=1}^{D}\sum_{i=1}^{n_j}\omega_{ij}^{(t)} -\sum_{j=1}^{D}\sum_{i=1}^{n_j} \omega_{ij}^{(t)} \rho_\tau\left(\frac{y_{ij}-\M{x}_{ij}^{
    \T} \widehat{\bm{\beta}}_{\tau}^{(t+1)}}{\sigma_{\tau}} \right)\right\}.
\end{eqnarray*}
We note that the update for $\bm{\beta}$ and $\sigma$ is an approximate M-step consisting of a single cycle of coordinate descent in these two parameters. The update of $\bm{\beta}$ can be approximated further via a weighted least squares fit \citep[e.g.,][$\S$4]{Yohai2006} associated with 
(predictors, response)-pairs $(\M{x}_{ij},y_{ij})$ and weights $c_{ij}^{(t)}=\rho_{\tau}'(r^{(t)}_{ij})/r^{(t)}_{ij}$ where $r^{(t)}_{ij}=(y_{ij}-\mathbf{x}_{ij}^{\T}\widehat{\bm{\beta}}^{(t)}_{\tau})/\widehat{\sigma}^{(t)}_{\tau}$, $i=1,\dots,n_j, \, j=1,\ldots,D$. 
Accordingly, we use the update
\begin{equation}\label{mq_beta_hat}
    \widehat{\bm{\beta}}_{\tau}^{(t+1)}=\left(\mathbf{X}^{\T} \mathbf{W}_{\tau}^{(t)} \mathbf{C}_{\tau}^{(t)} \mathbf{X} \right)^{-1}
    \mathbf{X}^{\T}\mathbf{W}_{\tau}^{(t)} \mathbf{C}_{\tau}^{(t)}\mathbf{y},
\end{equation}
where $\mathbf{X}$ is the $n$-by-$d$ matrix and $\mathbf{y}$ is the vector of length $n$ obtained stacking $\{ \mathbf{x}_{ij} \}$ and $\{ y_{ij} \}$ respectively, and $\mathbf{W}_{\tau}^{(t)}=\diag\big( (\omega_{ij}^{(t)}) \big)$ and $\mathbf{C}_{\tau}^{(t)} = \diag\big( (c_{ij}^{(t)}) \big)$ are diagonal weight matrices of size $n$-by-$n$. 

Given the updated regression coefficients, we use the following update for $\sigma$:
\begin{equation*}
  \widehat{\sigma}_{\tau}^{(t+1)}=\left\{ \left(\sum_{j=1}^{D}\sum_{i=1}^{n_j} \omega_{ij}^{(t)} \right)^{-1} \sum_{j=1}^{D}\sum_{i=1}^{n_j} \omega_{ij}^{(t)}c_{ij}^{(t)} \left(y_{ij} -\M{x}_{ij}^{
    \T} \widehat{\bm{\beta}}_{\tau}^{(t+1)}\right)^{2}\right\}^{1/2}  
\end{equation*}

Alternatively, $\sigma$ can be updated according to quantile-like generalization of the Median Absolute Deviation (MAD) estimator, i.e.
\begin{equation}\nonumber
\widehat{\sigma}_{\tau}^{(t+1)} 
= \frac{q_{\tau}\left( \{ | y_{ij} \, – \, \M{x}_{ij}^{\T} \wh{\bm{\beta}}_{\tau}^{(t)}| \}_{i=1, j=1}^{n_j, D} \right)}{\Phi((\tau + 1)/2)}
\end{equation}
where $q_{\tau}$ denotes the empirical $\tau$-quantile function and $\Phi$ is the standard Normal CDF. 

\subsection{MSE estimation}\label{sec:msemq}
{\bfseries Step 1: Variability in $\wh{\bm{\beta}}$.} An estimator of the MSE for \eqref{mq_predictor} can be based on the linearization approach proposed by \cite{Chambers2014}.  Assuming that $\tau_j$ and the $\{\bm\alpha_{\tau}: \, \tau \in T\}$ are known, the prediction error of $\widehat{\overline{y}}_{j}^{\text{MQ}}$ can be expressed as %
\begin{eqnarray}\label{MQ_pred_error}
\nonumber \widehat{\overline{y}}_{j}^{\text{MQ}}-\overline{y}_{j}  &= &\frac{1}{N_j} 
\sum_{ i = 1}^{N_j} \M{x}_{ij}^{\T} \wh{\bm{\beta}}_{{\tau}_{j}}-
\frac{1}{N_j} \sum_{ i = 1}^{N_j} y_{ij}
= -(\overline{y}_{j} -\overline{\M{x}}_j^{\T}\wh{\bm{\beta}}_{{\tau}_{j}}).
\end{eqnarray}
A first-order approximation to the MSE of $\widehat{\overline{y}}_{j}^{\text{MQ}}$ is then given by
\begin{eqnarray}\label{MQ_MSE_lin}
\text{MSE}\left(\widehat{\overline{y}}_{j}^{\text{MQ}} |\mc{D},\tau_j, \{ \bm{\alpha}_{\tau}\}_{\tau \in T} \right)&=&E\left[\left(\overline{y}_{j} -\overline{\M{x}}_j^{\T}\wh{\bm{\beta}}_{{\tau}_{j}}\right)^2 |\mc{D},\tau_j, \{ \bm{\alpha}_{\tau}\}_{\tau \in T} \right] \\
&+&\text{B}^2\left(\widehat{\overline{y}}_j^{\text{MQ}}|\mc{D},\tau_j, \{ \bm{\alpha}_{\tau}\}_{\tau \in T} \right)+o(n^{-2}),\nonumber
\end{eqnarray}
where $E\left[\left(\overline{y}_{j} -\overline{\M{x}}_j^{\T}\wh{\bm{\beta}}_{{\tau}_{j}}\right)^2|\mc{D},\tau_j, \{ \bm{\alpha}_{\tau}\}_{\tau \in T}\right] =\frac{1}{N_j(n_j-1)}\sum_{ i = 1}^{n_j} (y_{ij}^{(s)}-\M{x}_{ij}^{(s)\T} \wh{\bm{\beta}}_{{\tau}_{j}})^2$, 
and according to arguments in \citet{Chambers2014, chambers2011}, the bias term of $\widehat{\overline{y}}_j^{\text{MQ}}$ can be expressed as 
\begin{equation}\label{mq_mse_biascomp}
\text{B}\left(\widehat{\overline{y}}_j^{\text{MQ}}|\mc{D}, \{ \tau_{\ell} \}_{\ell = 1}^D, \{ \bm{\alpha}_{\tau}\}_{\tau \in T} \right) =
 \left\{\sum_{\ell = 1}^D \sum_{i = 1}^{n_{\ell}} b_{i{\ell}}^j \mathbf{x}_{i\ell}^{(s)\T} \widehat{\bm{\beta}}_{{\tau}_{\ell}} 
- \frac{1}{N_j}\sum_{i=1}^{N_j } \mathbf{x}_{ij}^{\T} \widehat{\bm{\beta}}_{{\tau}_{j}} \right\},
\end{equation}
where the $\{ b_{i\ell}^j \}$ are the elements of a vector $\mathbf{b}^j$ of length $n$ defined as
\[
\mathbf{b}^j=(b_{i\ell}^j) =
\mathbf{W}_{{\tau}_j}\mathbf{C}_{{\tau}_{j}} \mathbf{X}^{(s)}
\left(\mathbf{X}^{(s) \T}\mathbf{W}_{{\tau}_{j}}\mathbf{C}_{{\tau}_{j}}\mathbf{X}^{(s)} \right)^{-1}
\overline{\mathbf{x}}_{j}. 
\]
In the above equation, the diagonal weight matrices $\M{C}_{{{\tau}_{j}}}$ and $\M{W}_{{{\tau}_{j}}}$ are defined analogously as in the previous subsection (cf.~Eq.~\eqref{mq_beta_hat}). Note that the $\{ b_{i\ell}^j \}$ are non-negative and that  $\sum_{\ell = 1}^D \sum_{i=1}^{n_{\ell}} b_{i\ell}^j=1$; moreover, we assume that $b_{i\ell}^j = O(n_j^{-1})$ when $\ell=j$, 
and  $b_{i\ell}^{j} = o(n_j^{-1})$  for all $\ell \neq j$  \citep{chambers2011}.

\vskip1ex
{\bfseries Step 2: Variability in $\wh{\tau}_j$.} In practice, $\tau_j$ and the $\{ \bm{\alpha}_{\tau} \}_{\tau \in T}$ are unknown and need to be estimated from the data. Then an estimator of the MSE for $\widehat{\overline{y}}_{j}^{\text{MQ}}$ is obtained by starting from \eqref{MQ_MSE_lin}  with $\tau_j$ replaced by $\wh{\tau}_j$ and $\{ \bm{\alpha}_{\tau} \}_{\tau \in T}$ replaced by $\{ \wh{\bm{\alpha}}_{\tau} \}_{\tau \in T}$. To account for the estimation of $\tau_j$ and $\{ \bm{\alpha}_{\tau} \}_{\tau \in T}$ the prediction error of $\widehat{\overline{y}}_{j}^{\text{MQ}}$ may be decomposed as
\begin{equation*}
    \widehat{\overline{y}}_{j}^{\text{MQ}} -\overline{y}_{j}=\left[\widetilde{\overline{y}}_{j}^{\text{MQ}}-\overline{y}_{j}\right]+\left[\widehat{\overline{y}}_{j}^{\text{MQ}}-\widetilde{\overline{y}}_{j}^{\text{MQ}}\right],
\end{equation*}
where $\widetilde{\overline{y}}_j^{\text{MQ}}$ here denotes the M-quantile based predictor with known $\tau_j$ and $\{ \bm{\alpha}_{\tau}\}_{\tau \in T}$. Taking the expectation of the squared difference, we obtain
\begin{equation*}  
\text{MSE}(\widehat{\overline{y}}_{j}^{\text{MQ}})=\text{MSE}(\widetilde{\overline{y}}_{j}^{\text{MQ}})+\E\left[(\widehat{\overline{y}}_{j}^{\text{MQ}}-\widetilde{\overline{y}}_{j}^{\text{MQ}})^2 \right]+2 \E[(\widetilde{\overline{y}}_{j}^{\text{MQ}}-\overline{y}_{j}) (\widehat{\overline{y}}_{j}^{\text{MQ}}-\widetilde{\overline{y}}_{j}^{\text{MQ}})],
\end{equation*}
where $\text{MSE}(\widetilde{\overline{y}}_{j}^{\text{MQ}})$ can be computed according to \eqref{MQ_MSE_lin} and \eqref{mq_mse_biascomp}. The second and the third term above are generally not tractable, and are hence approximated using a Taylor expansion. \citet{Bertarelli:2024} provide the order of the negligible terms in these approximations.

By a first-order Taylor expansion of $\widehat{\overline{y}}_j^{\text{MQ}}$ around $\tau_j$, we obtain
\begin{equation}\label{delta_MQC}
\widehat{\overline{y}}_j^{\text{MQ}}-\widetilde{\overline{y}}_j^{\text{MQ}}=\delta(\tau_j)(\widehat{\tau}_j-\tau_j)+o(n_L^{-1}),
\end{equation}
where $\delta(\tau_j)=\partial{\widehat{\overline{y}}_j^{\text{MQ}}}/\partial \tau_j$ and $n_L \coloneq \min_{1 \leq \ell \leq D} n_{\ell}$. Moreover, we assume that terms involving higher powers of $\wh{\tau}_j-\tau_j$ are of lower order than $\delta(\tau_j)(\wh{\tau}_j-\tau_j)$.
We then have
\begin{equation*}
\delta(\tau_j) =\overline{\M{x}}_j^{\T}\frac{\partial \widehat{\bm{\beta}}_{{\tau}_{j}} }{\partial \tau_j} \invcoloneq \triangle(\tau_j).
\end{equation*}
It now follows that
\begin{equation}\label{delta_MQC2}
\E\left[(\widehat{\overline{y}}_j^{\text{MQ}}-\widetilde{\overline{y}}_j^{\text{MQ}})^2 \right] = \E\left[\{ \triangle(\tau_j)(\widehat{\tau}_j-\tau_j) \}^2 \right]+o(n_L^{-2}),
\end{equation}
with 
\begin{equation*}
    \E\left[\{ \triangle(\tau_j)(\widehat{\tau}_j-\tau_j) \}^2 \right] \approx  \var(\widehat{\tau}_j)\triangle^2(\tau_j) \invcoloneq V(\wh{\tau}_j).
\end{equation*}

{\bfseries Variability in $\{ \wh{\bm{\alpha}}_{\tau} \}_{\tau \in T}$}. We can extend the above approach to incorporate the uncertainty resulting from the estimation of the parameters $\{ \bm{\alpha}_{\tau} \}_{\tau \in T}$, although explicit formulas will depend on the specification of the function $h(\cdot;\bm{\alpha}_{\tau})$. By a first-order Taylor expansion of $\widehat{\overline{y}}_j^{\text{MQ}}$ around $\bm{\alpha}$, we obtain
\begin{equation}\label{delta_MQC_alpha}
\widehat{\overline{y}}_j^{\text{MQ}}-\widetilde{\overline{y}}_j^{\text{MQ}}=\delta\left(\{ \bm{\alpha}_{\tau} \}_{\tau \in T}\right)\left(\{ \widehat{\bm{\alpha}}_{\tau} \}_{\tau \in T}-\{ \bm{\alpha}_{\tau} \}_{\tau \in T}\right)+o(n_L^{-1}),
\end{equation}
where 
\begin{equation*}
\delta\left(\{ \bm{\alpha}_{\tau} \}_{\tau \in T}\right) =\overline{\M{x}}_j^{\T}\frac{\partial \widehat{\bm{\beta}}_{{\tau}_{j}} }{\partial \{ \bm{\alpha}_{\tau} \}_{\tau \in T}} \invcoloneq \bm{\Delta}(\{ \bm{\alpha}_{\tau} \}_{\tau \in T}).
\end{equation*}
Therefore, 
\begin{equation*}
  \E\left[(\widehat{\overline{y}}_j^{\text{MQ}}-\widetilde{\overline{y}}_j^{\text{MQ}})^2 \right]=  \bm{\Delta}\left(\{\bm{\alpha}_{\tau} \}_{\tau \in T}\right)\cov\left(\{\widehat{\bm{\alpha}}_{\tau} \}_{\tau \in T}\right) \bm{\Delta}^{\T}\left(\{\bm{\alpha}_{\tau} \}_{\tau \in T}\right) \invcoloneq V\left(\{\wh{\bm{\alpha}}_{\tau} \}_{\tau \in T}\right).
\end{equation*}
An estimator of the MSE of $\widehat{\overline{y}}_j^{\text{MQ}}$ can be obtained along the lines of \cite{bianchi2015asymptotic}, as considered also in \cite{salvati2021}. In combination, we have 
\begin{eqnarray}\label{MSE2}
\text{MSE}\left(\widehat{\overline{y}}_{j}^{\text{MQ}}\right)=\frac{1}{N_j (n_j-1)}\sum_{ i = 1}^{n_j} (y_{ij}^{(s)}-\M{x}_{ij}^{(s)\T} \wh{\bm{\beta}}_{{\tau}_{j}})^2+{B}^2\left(\widehat{\overline{y}}_j^{\text{MQ}}\right)+V\left(\widehat{\tau}_{j}\right) +V\left(\{\wh{\bm{\alpha}}_{\tau} \}_{\tau \in T}  \right) +o(n^{-2}),
\end{eqnarray}
and an estimator of \eqref{MSE2} is given by
\begin{eqnarray}\label{MSE3}
\wh{\text{MSE}}\left(\widehat{\overline{y}}_{j}^{\text{MQ}}\right)\approx \frac{1}{N_j (n_j-1)}\sum_{ i = 1}^{n_j} (y_{ij}^{(s)}-\M{x}_{ij}^{(s)\T} \wh{\bm{\beta}}_{\wh{\tau}_{j}})^2+\wh{B}^2\left(\widehat{\overline{y}}_j^{\text{MQ}}\right)+\wh{V}\left(\widehat{\tau}_{j}\right) +\wh{V}\left(\{\wh{\bm{\alpha}}_{\tau} \}_{\tau \in T}  \right),
\end{eqnarray}
where
\begin{align*}
&\wh{B}\left(\widehat{\overline{y}}_j^{\text{MQ}} \right) = \left\{\sum_{\ell = 1}^D \sum_{i = 1}^{n_{\ell}} b_{i{\ell}}^j \mathbf{x}_{i\ell}^{(s)\T} \widehat{\bm{\beta}}_{{\wh{\tau}}_{\ell}} 
- \frac{1}{N_j}\sum_{i=1}^{N_j } \mathbf{x}_{ij}^{\T} \widehat{\bm{\beta}}_{{\wh{\tau}}_{j}} \right\},
\quad
\wh{V}\left(\widehat{\tau}_{j}\right)  = (n_j-1)^{-1}\sum_{i=1}^{n_j}\left(\widehat{\tau}_{ij}- \widehat{\tau}_{j} \right)^2 \triangle^2(\widehat{\tau}_j),\\
&\mbox{{\small $\wh{V}\left(\{\wh{\bm{\alpha}}_{\tau} \}_{\tau \in T}  \right) = \bm{\Delta}\left(\{\wh{\bm{\alpha}}_{\tau} \}_{\tau \in T}\right)  \bigg( \displaystyle\sum_{j = 1}^{D} \sum_{i=1}^{n_j} \frac{\partial}{\partial \{ \boldsymbol{\alpha}_{\tau} \}_{\tau \in T}} k_{ij} \big(\{\wh{\boldsymbol{\alpha}}_{\tau}\}_{\tau \in T} \big) \bigg)^{-1} \displaystyle\sum_{j = 1}^{D} \displaystyle\sum_{i=1}^{n_j} k_{ij} \big(\{ \wh{\boldsymbol{\alpha}}_{\tau} \}_{\tau \in T} \big)  k_{ij}\big(\{ \wh{\boldsymbol{\alpha}}_{\tau} \}_{\tau \in T}\big)^{\T}$}}\\
&\qquad \qquad \qquad \qquad \qquad \quad \mbox{{\small $\left \{ \bigg( \displaystyle\sum_{j = 1}^{D} \sum_{i=1}^{n_j} \frac{\partial}{\partial \{ \boldsymbol{\alpha}_{\tau} \}_{\tau \in T}} k_{ij}\big( \{ \wh{\boldsymbol{\alpha}}_{\tau} \}_{\tau \in T} \big) \bigg)^{-1} \right\}^{\T}\bm{\Delta}^{\T}\left(\{\wh{\bm{\alpha}}_{\tau} \}_{\tau \in T}\right)$}},
\end{align*}
with $k_{ij}(\{ \wh{\boldsymbol{\alpha}}_{\tau} \}_{\tau \in T})=\bigg( \nabla_{\bm{\alpha}} h(\M{d}_{ij}^{(s)};{\wh{\boldsymbol{\alpha}}}_{\tau})
\left\{  
\frac{\omega_{ij}}{1-h(\M{d}_{ij}^{(s)};{\wh{\boldsymbol{\alpha}}}_{\tau})} - 
\frac{1-\omega_{ij}}{h(\M{d}_{ij}^{(s)};{\wh{\boldsymbol{\alpha}}_{\tau}})}
\right\} \bigg)_{\tau \in T}$.

\section{Empirical evaluation}\label{simulations}
In this section, we present the results of simulations to showcase the empirical performance of the 
two approaches proposed above. We consider both a model-based and a design-based simulation. In the former,  
data are generated that precisely follow the linear mixed effects SA model \eqref{eq:unitlevel_SA} with 
mismatch error introduced uniformly at random according to a procedure described below in a way that guarantees 
 that the assumptions in $\S$\ref{assumptions} are
satisfied. We also consider a modification that includes heteroscedastic random effects and the presence of outliers. In the design-based simulation, only the sampling of units and the record linkage process are generated synthetically. The specifics of the two simulations are adopted from \cite{salvati2021}. In the sequel, we review these specifics and how the approaches proposed herein are run in each case. Moreover, we present comparisons to several baseline competitors.  

\subsection{Model-based simulation}
In the model-based simulations, the synthetic finite population consists of $D = 40$ areas with population sizes 
$N_j = 100$, $1 \leq j \leq D$. Data are generated according to the linear mixed effect models \eqref{eq:unitlevel_SA} with a single predictor variable and area-specific intercepts. Specifically, 
\begin{align}\label{eq:mbased_0_0}
&x_{ij} \overset{\text{i.i.d.}}{\sim} \exp(N(1, 0.25)), \quad \gamma_j \overset{\text{i.i.d.}}{\sim} N(0, 6),  \quad \epsilon_{ij} \overset{\text{i.i.d.}}{\sim} N(0, 3), \notag\\
&y_{ij} = 100 + 5 x_{ij} + \gamma_j + \epsilon_{ij}, \quad 1 \leq i \leq N_j, \; 1 \leq j \leq D, 
\end{align}
where $\exp(N(\ldots))$ is a shorthand for the log-Normal distribution.  Linkage between 
the $\{ x_{ij} \}$ and the $\{ y_{ij} \}$ is mimicked at the population level, area-by-area (thus excluding
linkage error across areas, in line with \cite{salvati2021} even if this is not an assumption of the methods discussed herein). For each 
area $1 \leq j \leq D$, a random permutation matrix $\bm\Pi_j^{\star}$ is generated so that its expectation aligns with the exchangeable linkage error model \citep{Chambers2009, salvati2021}. Specifically, the number of blocks 
is given by four (each of which are of equal size $N_j/4 = 25$), with mismatch rates equal to 
$\alpha_1^* = 0$, $\alpha_2^* = 0.1$, $\alpha_3^* = 0.4$, and $\alpha_4^* = 0.6$. Given linked pairs (some of which are incorrectly matched), $n_j = 5$ units are sampled for each area uniformly at random, yielding a data set $\{ x_{ij}^{(s)}, z_{ij}^{(s)} = 1, y_{ij}^{(s)}\}_{i = 1, j = 1}^{5, 40}$. 

We also perform simulations for the following departures from \eqref{eq:mbased_0_0}: (i) for 10\ of the areas 
(37 through 40), the random effects $\{ \gamma_j \}_{j = 37}^{40}$ are drawn i.i.d.~from the $N(0,20)$ distribution, 
and (ii) the errors $\{ \epsilon_{ij} \}$ are drawn from a $\delta$-contaminated Gaussian distribution $(1-\delta) N(0,3) + \delta N(0, 150)$, where $\delta = 0.03$ is the rate of ``outliers" injected into the sample. When reporting results, we refer to \eqref{eq:mbased_0_0} as setting \textsf{(0,0)} and to the setting including the departures (i) and (ii)
as  \textsf{(0.1,0.03)}. 

The proposed approaches are run under three different scenarios: (i) the mismatch rates 
and blocks are known, in which case we use $h(\M{d}_{ij}^{(s)};\bm{\alpha}) = \sum_{q = 1}^4 \alpha_q \mathbb{I}\{ (i,j) \in \text{block}_q \}$, (ii) the blocks are known but the mismatch rates are unknown, in which case we use $h(\M{d}_{ij}^{(s)};\bm{\alpha}) = \sum_{q = 1}^4 \alpha_q \mathbb{I}\{ (i,j) \in \text{block}_q \}$ with the $\{ \alpha_q \}_{q = 1}^4$ to be estimated from the data, and (iii) neither mismatch rates nor blocks are known in 
which case we use $h(\M{d}_{ij}^{(s)};\overline{\alpha}) = \overline{\alpha}$ for a global 
mismatch rate $\overline{\alpha}$ to be estimated from the data, $1 \leq i \leq n_j$, $1 \leq j \leq D$. 

The approach based on the linear mixed model is run with the density $g$ estimated 
by a kernel density estimator with Gaussian kernel and bandwidth chosen according to Silverman's rule of thumb \citep{Silverman2018}. The EM iterations are initialized 
with initial iterates $\{ \bm{\beta}^{(0)}, \Sigma^{(0)}, \sigma^{2(0)} \}$ obtained from naive REML estimation without adjustment for the presence of mismatches. The initial
iterates for the mismatch rates are all set to $0.1$ in both scenarios (ii) and (iii). The EM algorithm is run for a fixed number of iterations ($200$) with exact computations in the E-step given small sample sizes within each area. Estimates of the MSE are obtained as described in $\S$\ref{sec:MSEEBLUP}: first, quantities appearing in Louis' formula \eqref{eq:covariance_louis} are approximated by $10^4$ Monte-Carlo samples, which are then substituted into \eqref{eq:asymptotic_composite} to generate $M = 100$ samples $\{ \wt{\bm{\beta}}^{[m]} \}_{m = 1}^M$ under scenario (i) and additionally $\{ \wt{\bm{\alpha}}^{[m]} \}_{m = 1}^M$ under scenarios (ii) and (iii) to be used in \eqref{eq:MSE_MC} with $\Sigma_*$ and $\sigma_*^2$ replaced by the estimates obtained from our approach. 

The M-quantile based approach is run as follows:
the EM iterations are initialized with $\{\widehat{\bm{\beta}}_{\tau}^{(0)}, \wh{\sigma}_{\tau}^{(0)}\}$
obtained from naive M-quantile estimation without adjustment for the presence of mismatches. The initial iterates for the mismatch rates are all set to 0.1 in both scenarios (ii) and (iii). The EM algorithm is run for a fixed number of iterations (100). The observation-specific quantiles $\{ \widehat{\tau}_{ij} \}$ are obtained by computing $\{ \widehat{\bm{\beta}}_\tau \}_{\tau \in T}$ for a grid of quantiles $T = \{0.02, 0.05, \ldots, 0.98\}$ and then solving the equation $y_{ij}=\M{x}_{ij}^{\T}\widehat{\bm{\beta}}_{\tau}$ in $\tau$ by linear interpolation. The loss function $\rho$ is chosen as Huber's loss function \citep[e.g.,][Eq.~(2.28)]{Yohai2006} with tuning constant equal to $1.345$. The MSE is estimated using expression \eqref{MQ_MSE_lin}. 
\begin{table}[h!!!!!!!!!!!!!!!!!!!!!!!!!!!!!!!!!]
    \centering
    {\small \begin{tabular}{|l|rrrrr|rrrrr|}\hline
       & \multicolumn{5}{|c}{\textsf{(0,0)}} & \multicolumn{5}{|c|}{\textsf{(0.1, 0.03)}} \\[-.55ex]
       & \multicolumn{5}{|c|}{Parameter} & \multicolumn{5}{|c|}{Parameter} \\[-.55ex] 
       Predictor  & $\beta_0^*$ &  $\beta_1^*$ & $\alpha_*$ & $\Sigma_*$ & $\sigma_*^2$ & $\beta_0^*$ &  $\beta_1^*$ & $\alpha_*$ & $\Sigma_*$ & $\sigma_*^2$ \\[-.25ex] \hline
      $^{1}EBLUP$   & 4.2 & -27.3 & & -2.4 & 534.2 & 4.3 & -27.8 & & 52.2 & 600.9\\[-.55ex]
      $^{2}EBLUP^{\star}$ & 0.0 & -0.1 & & 0.2 & 330.1 & 0.0 & 0.1 & & 53.1 & 398.8 \\[-.55ex]
      $^{3}EBLUPME~ab$ & 0.1 & -0.3 & & -8.0 & 2.5  & 0.0 & -0.2 & & 50.0 & 20.3 \\[-.55ex]
      $^{4}EBLUPME~b$ & 0.1 & -0.3 & -0.9 &-8.1 & 1.8 & 0.0 & 0.2& 6.5 & 47.4 & 0.6 \\[-.55ex]
      $^{5}EBLUPME$  & 0.1 & -0.4 & -1.3 &-6.4 & 1.6 & 0.0 & 0.0 & 6.2 & 49.5 & -0.1\\[-.55ex]
      $^{6}MQ$  & 2.0 & -13.1 & & &  & 2.1 & -14.0& & & \\[-.55ex]
      $^{7}MQ^{\star}$  & 0.0 & -0.1 & & &  & 0.0 &-0.1 & & & \\[-.55ex]
      $^{8}MQME~ab$  & 0.0 & -0.4 & &  & & 0.0 & -0.2& & & \\[-.55ex]
      $^{9}MQMQ~b$  & 0.0 & -0.6 & -11.9& &  & 0.0 & -0.1 & -6.2 & &\\[-.55ex]
      $^{10}MQMQ$  & 0.0 & -0.9 & -13.0& &  & 0.0 & -0.5 & -7.2 & &\\ \hline
    \end{tabular}}
    \vskip1ex
    {\footnotesize \begin{tabular}{|l|l||l|l|}
    \hline
      & Explanation &  & Explanation \\
     \hline
     $1$ & EBLUP w/o adjustment for linkage error & $6$ & MQ w/o adjustment for linkage error\\
     $2$ & EBLUP adjustment in \cite{salvati2021} & $7$ & MQ adjustment in \cite{salvati2021} \\
     $3$ & proposed EBLUP adjustment: $\alpha$'s + blocks known & $8$  & proposed MQ adjustment: $\alpha$'s + blocks known \\
      $4$   & proposed EBLUP adjustment: blocks known        & $8$ & proposed MQ adjustment:  blocks known \\
      $5$ & proposed EBLUP adjustment (plain) & $10$ & proposed MQ adjustment (plain)\\ 
      \hline
    \end{tabular}}
    \caption{Top: Relative bias (as percentage) of the estimators under consideration with regard to the model parameters under scenarios \textsf{(0, 0)} and \textsf{(0.1, 0.03)}. Here, $\alpha_*$ refers to the average mismatch rates over all four blocks. Bottom: Legend listing the approaches that are compared.}
    \label{tab:bias}
\end{table}

When reporting the results, references to specific methods are made as follows. Under scenario (i) the linear mixed model-based predictors and the MQ-based predictor are referred to as $EBLUPME~ab$ and $MQME~ab$, respectively; in scenario (ii) the two predictors are referred to as $EBLUPME~b$ and $MQME~b$; finally, under the last scenario where neither mismatch rates nor blocks are known, the estimators are referred to as $EBLUPME$ and $MQME$. The following baselines are run for the sake of comparison: the standard EBLUP \citep{rao2015}, the EBLUP adjustment for linkage error in \cite{salvati2021} (denoted by $EBLUP^{\star}$), the estimator based on the MQ regression model \citep{Chambers2006} and the adjustment of this method for linkage error developed in \cite{salvati2021}. Moreover, the results for the EBLUP and MQ-based predictor in the absence of mismatches (referred to as ``oracles") are reported to evaluate the loss of the proposed predictors in terms of bias and efficiency in case of mismatches. 

\begin{figure}
    \centering
    \includegraphics[scale=0.5]{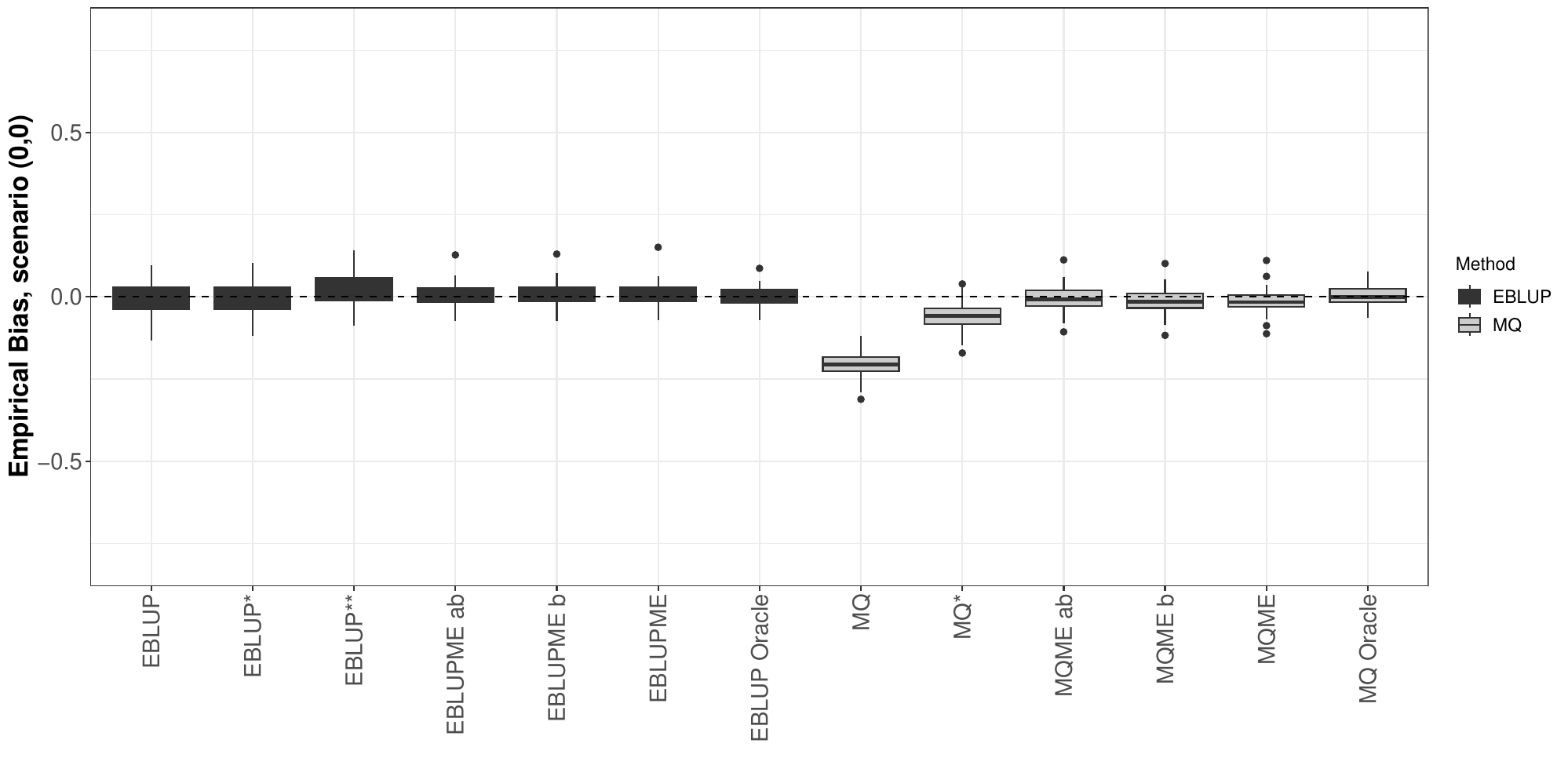}\\
    \includegraphics[scale=0.5]{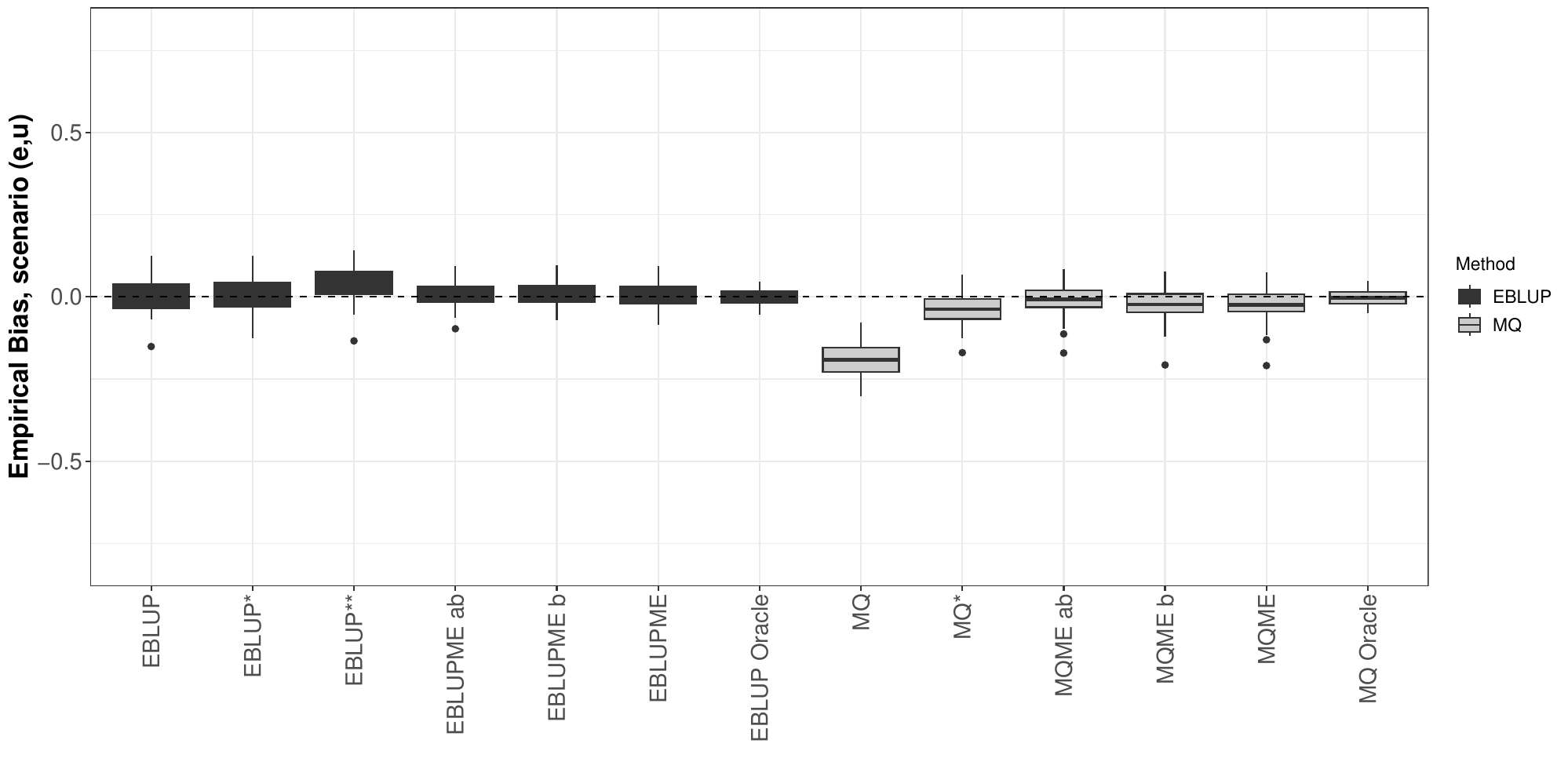}
    \caption{Boxplots showing area-specific values of the empirical bias of the small area estimators in the model-based experiment under the scenario \textsf{(0,0)} (top panel) and scenario \textsf{(0.1,0.03)} (bottom panel).}
    \label{fig:mb:bias}
\end{figure}

\begin{figure}
    \centering
    \includegraphics[scale=0.5]{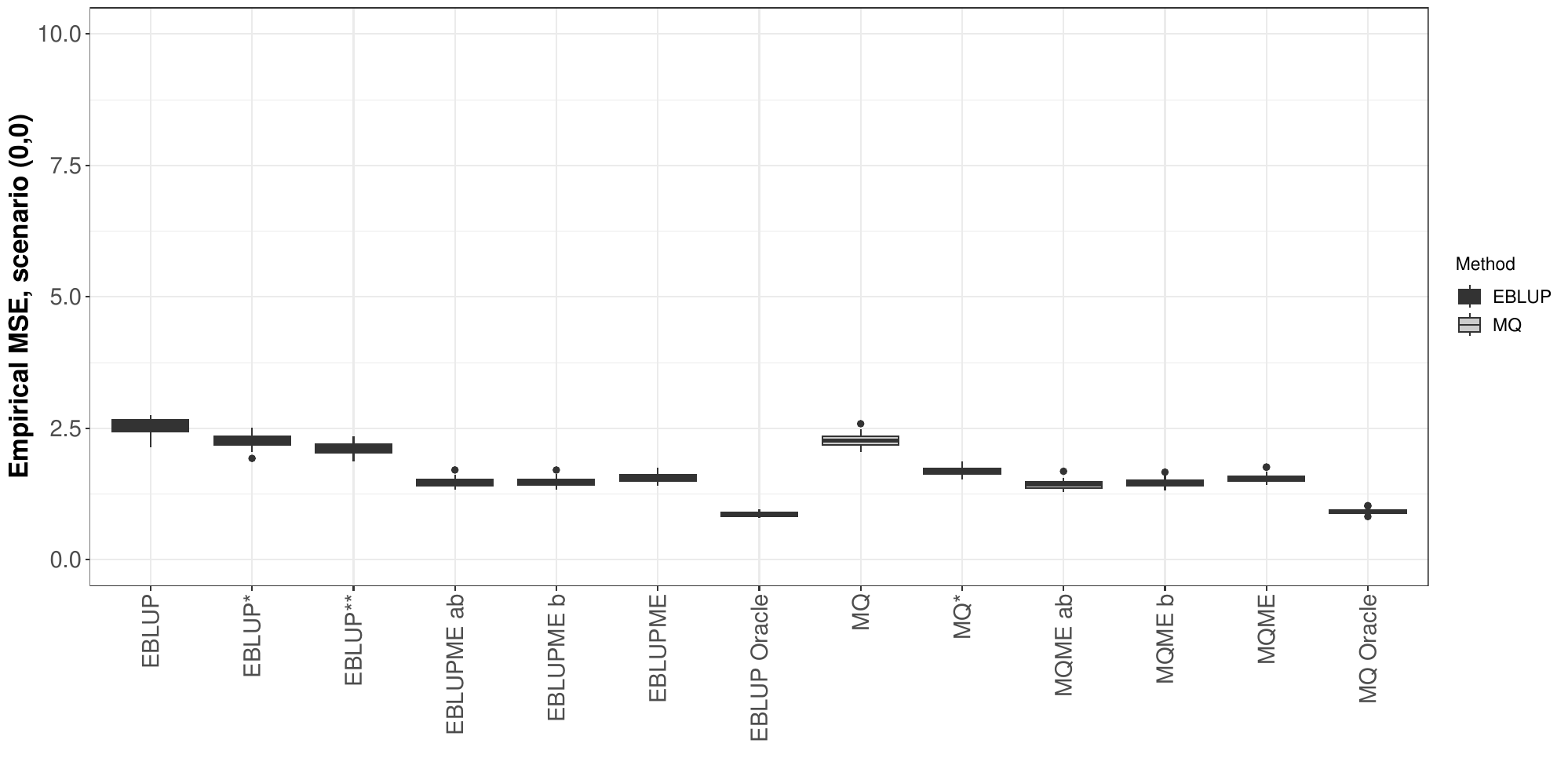}\\
    \includegraphics[scale=0.5]{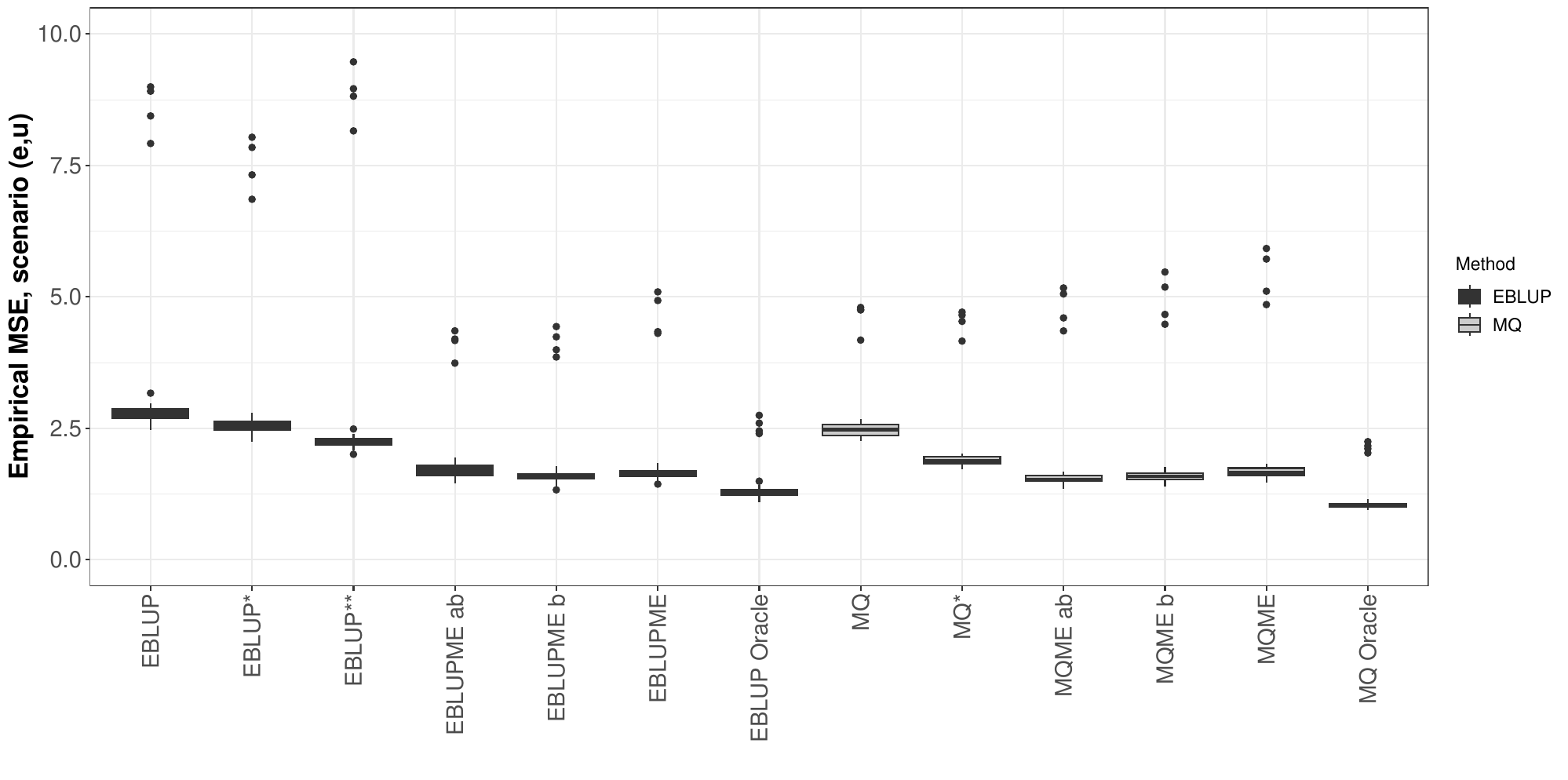}
    \caption{Boxplots showing area-specific values of the empirical MSE of the small area estimators in the model-based experiment under the scenario \textsf{(0,0)} (top panel) and scenario \textsf{(0.1,0.03)} (bottom panel).}
    \label{fig:mb:mse}
\end{figure}

\begin{figure}
    \centering
    \includegraphics[scale=0.25]{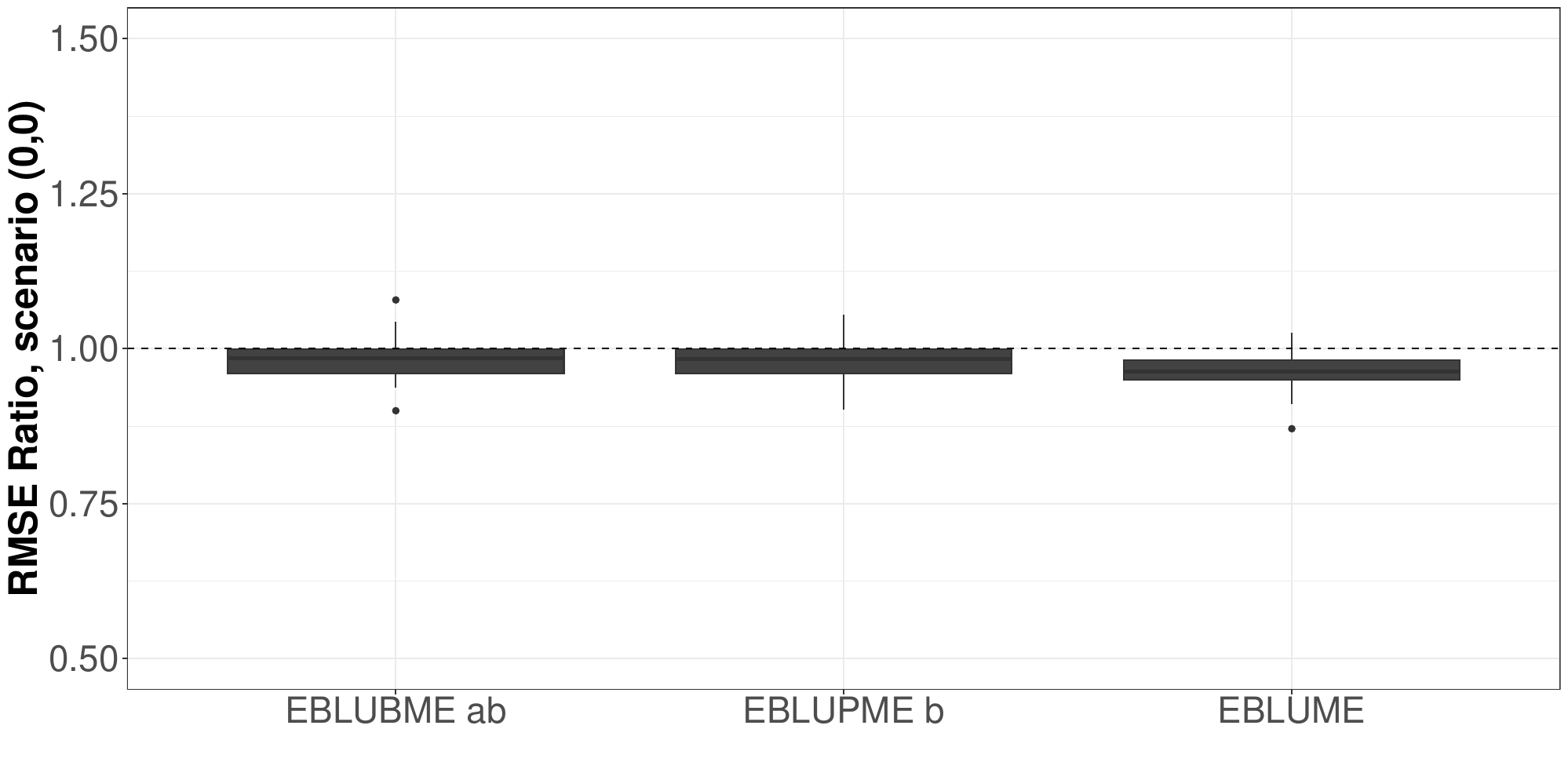}
    \includegraphics[scale=0.25]{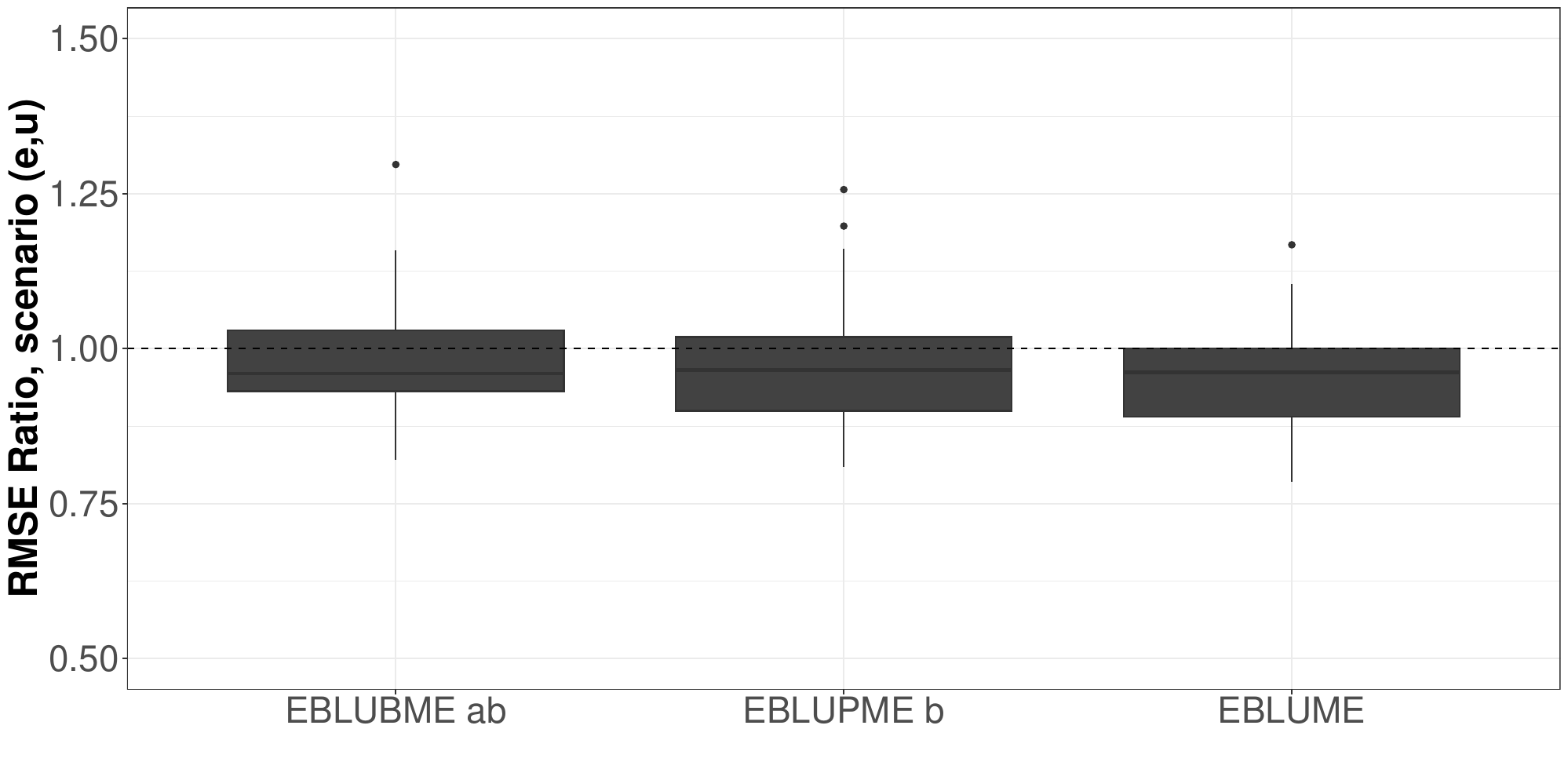}\\
    \includegraphics[scale=0.25]{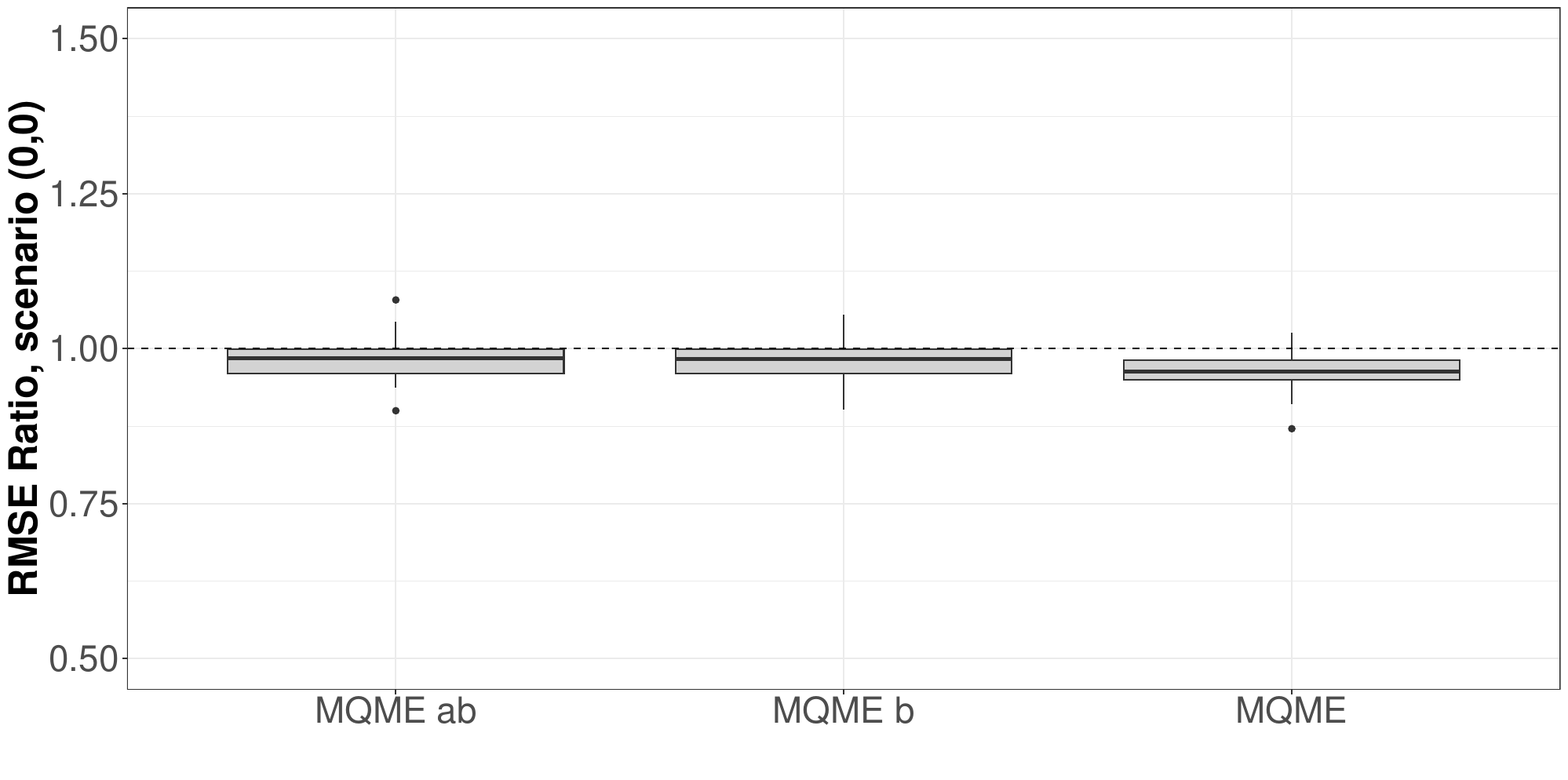}
    \includegraphics[scale=0.25]{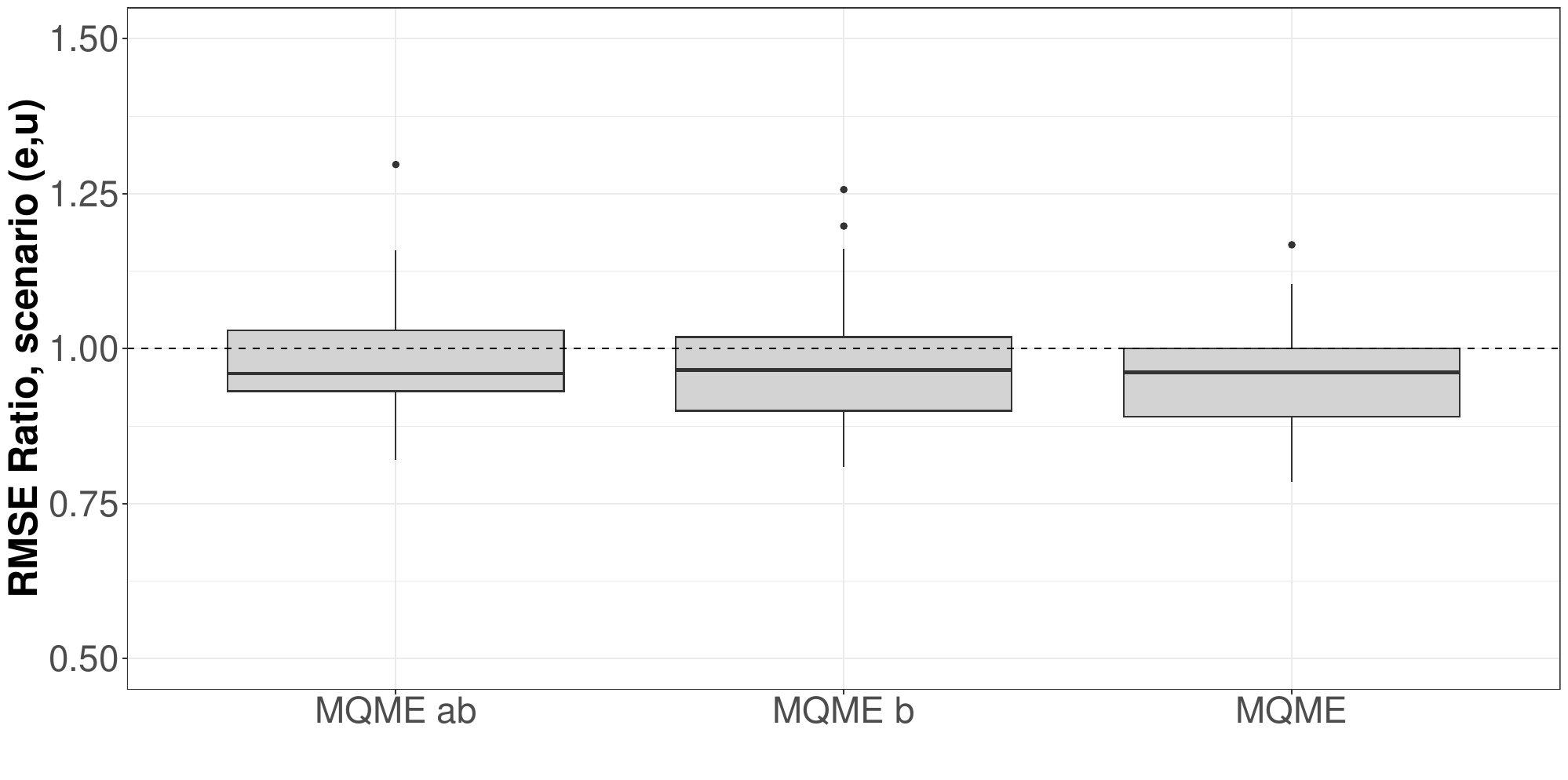}
    \caption{Boxplots showing area-specific values of the RMSE ratios of the MSE estimators in the model-based experiment. The RMSE ratio is defined as the ratio of the average over repeated sampling of the RMSE estimator for an estimator to the actual RMSE of this estimator under repeated sampling. Four graphs correspond to the two simulation scenarios and the EBLUP and MQ-based predictors; left panels (top for EBLUP and bottom for MQ) refer to the case \textsf{(0,0)}, whereas right panels (top for EBLUP and bottom for MQ) refer to the case \textsf{(0.1, 0.03)}.}
    \label{fig:mb:estMSE}
\end{figure}


Table \ref{tab:bias} shows the percentage relative bias of the form $100 \cdot (\overline{\wh{\theta}} - \theta_*) / \theta_*$ with regard to the estimation of the underlying parameters. In estimating regression coefficients, neither models with adjustment for the presence of mismatches exhibit any prevailing sign for the bias under all simulation scenarios. As expected the linear mixed effects and the M-quantile models without adjustment produce biased estimates of the regression coefficients. The two methods differ substantially in estimating $\alpha$. The M-quantile models exhibit downward bias (around $12\%$ to $13\%$ in scenario \textsf{(0,0)} and $6\%$ to $7\%$ in scenario \textsf{(0.1,0.03)}), whereas the estimates obtained with the proposed linear mixed models exhibit lower bias (around $-1\%$ in scenario \textsf{(0,0)} and $6\%$ in setting \textsf{(0.1,0.03)}). A likely explanation is that M-quantile regression better accommodates outliers and therefore tends to identify a smaller fraction of wrong links as such.
Moreover, the relative bias results that are set out in Table \ref{tab:bias} show a high level of bias for the estimators of the parameter $\sigma_*^2$ for the linear mixed effects models without adjustment for the presence of mismatches as well as with the correction suggested by \citet{salvati2021}. This increase in bias is most pronounced when there are outliers at individual and area levels. As expected, under the latter also the estimation of $\Sigma_*$ is affected, with all of the investigated methods exhibiting upward bias. This result suggests a robustification of the estimating equations of the linear mixed effects models as proposed in \citet{salvati2021} as a possible direction for future research.

Next, we examine the performance of the various small area predictors. For each estimator and for each small area, we computed the Monte Carlo estimate of the empirical bias and the empirical MSE. For a given area, the bias of an estimator is defined as the value of the average difference between the estimate and the corresponding true simulated small area mean, average being taken over simulations. The MSE is defined as the average squared difference between the estimate and the corresponding true simulated small area mean, average being taken over simulations.

Figures \ref{fig:mb:bias} and \ref{fig:mb:mse} present the distribution of the area-specific values of the empirical bias and MSE of the small area estimators under the two settings. In scenario \textsf{(0,0)} we see that the proposed estimators $EBLUPME~ab$, $MQME~ab$, $EBLUPME~b$, $MQME~b$, $EBLUPME$ and $MQME$ that allow for linkage error work well in terms of both bias and MSE compared with the $EBLUP^{\star}$, $MQ^{\star}$ and unmodified EBLUP and M-quantile-based predictors that ignore linkage error. The former are less biased and variable than the latter. Under scenario \textsf{(0.1,0.03)} the $EBLUPME~b$ and the $MQME~ab$ estimators perform best in terms of bias, whereas $EBLUPME~ab$, $MQME~ab$ perform best in terms of MSE. These results are expected because $EBLUPME~ab$ and $MQME~ab$ have the advantage of being equipped with the mismatch rates for each block. 

It is worth emphasizing that the predictors $EBLUPME$ and $MQME$, which correct for linkage error when neither mismatch rates nor blocks are known, achieve a performance both in terms of bias and MSE that is roughly on par with estimators correcting for linkage errors using a richer set of information. In particular, we note that $EBLUPME~ab$ exhibits a lower MSE than $EBLUP^{\star}$ and to a lesser extent the same is true for $MQ~ab$ with respect to $MQ^{\star}$. When the same set of information is available, the mixture specification proposed in this paper seems to be more effective in dealing with both representative and mismatch-induced outliers than their counterparts proposed previously in \cite{salvati2021}. In this context, according to a connection drawn in \cite{{slawski2021pseudo}}, we note the relationship between 
the two-component mixture model adopted for the approaches in this paper and the minimization of a ``capped" loss function resembling Tukey's celebrated biweight loss function. The latter is well known for its effectiveness in reducing the influence of large errors \citep[e.g.,][p.~30]{Yohai2006}.  


With regard to MSE estimation, the boxplots in Figure \ref{fig:mb:estMSE} are used to explore the variability in the RMSE ratio defined as the ratio of the average estimated RMSE for each area to the true RMSE. The left panels refer to the case \textsf{(0,0)}, whereas the right panels present the ratios for the case \textsf{(0.1,0.03)}. It can be seen that the MSE estimators of the two proposed approaches perform well especially for the MQ-based predictors.

The performance of the various small area predictors accounting for linkage error investigated in this study suggest that they are robust in the presence of such error as well as when additionally outliers are present. 

\subsection{Design-based simulation}

In this section, we adopt the simulation setup in \citet{salvati2021} to evaluate the performance of the estimators proposed in the case of a finite, fixed population and a realistic sampling method. The population we consider is synthetic but based on real data. It is derived from the European Statistical System Data Integration project \citep[ESSnet, cf.][]{McLeod2011} and the Survey on Household Income and Wealth, Bank of Italy (SHIW), which provides openly available anonymized data. Specifically, the synthetic ESSnet population includes information on over 26,000 individuals, encompassing variables such as first and last name, sex, and date of birth. Additionally, two supplementary variables, annual income, and domain indicators have been incorporated. The domain indicators represent 18 areas formed through the aggregation of Italian administrative regions, with population sizes ranging from 102 to 3262 individuals and an average of 1407. The setup of our simulation study is designed in close similarity to that in \citet{briscolini2018new}.

We conduct a realistic record linkage and Small Area Estimation (SAE) simulation experiment by perturbing the original ESSnet dataset. These perturbations involve the introduction of missing values and typographical errors in potential linking variables such as name, surname, gender, and date of birth. Furthermore, for the purpose of this simulation study, annual consumption obtained from the SHIW survey is also considered and will be used as the target variable in estimation.

We employ the classical probabilistic record linkage procedure proposed by \citet{Fellegi69} and \citet{Jaro1989}. This involves utilizing the \texttt{compare.linkage} function from the \linebreak \texttt{RecordLinkage} package in \texttt{R} \citep{Sariyar2020} to link the perturbed dataset with the original register population. The linkage is performed using surname as the key variable, with age categorized into four groups and domain serving as blocking variables. This implies that mismatch error only occurs within but not across domains, in line with the setting in \cite{salvati2021}. 




The linking process is done sample-to-register so that every sampled unit is linked to precisely one unit in the register. Although the linking is one-to-one, it is evidently incomplete and thus not perfectly in line with the assumption of complete linkage made in $\S$\ref{assumptions}. Empirically it turns out that the proposed predictors are not sensitive to this departure. Given that the linkage in this scenario is contingent on the realized sample, the proportion of correct links for the four age categories varies from one sample to another \citep{salvati2021}. 


The objective of the design-based simulation is to assess and contrast the effectiveness of various estimators along with their estimated MSEs in the absence of any distributional assumptions regarding the data-generating process, with annual mean consumption per domain serving as the target of interest ($y$). This assessment is conducted through repeated sampling from a synthetic fixed population, using annual income as the auxiliary variable ($x$). A total of 1,000 independent random samples, each consisting of 268 units, are drawn from the synthetic fixed population mentioned earlier. These samples are generated by randomly selecting units across the 18 domains, with the sample sizes within each domain being proportional to their respective population sizes. However, if the resulting sample size is less than 5, the domain sample size is fixed at 5.

The same estimators as in the model-based simulation are compared. They are run with the same configurations as described in the previous subsection. For each estimator and for each small area, we computed the Monte Carlo estimate of the relative bias (RB) and the relative root MSE (RRMSE). The relative bias of an estimator $\widehat{\overline{y}}_{j}$ of the actual mean ${\overline{y}}_{j}$ is the average across simulations of the errors $\widehat{\overline{y}}_{j}-{\overline{y}}_{j}$ divided by the corresponding value of ${\overline{y}}_{j}$ and its RRMSE is the square root of the average across simulations of the squares of these errors, again divided by the value of ${\overline{y}}_{j}$, $1 \leq j \leq D = 18$.

\begin{figure}[hb!]
    \centering
    \includegraphics[scale=0.53]{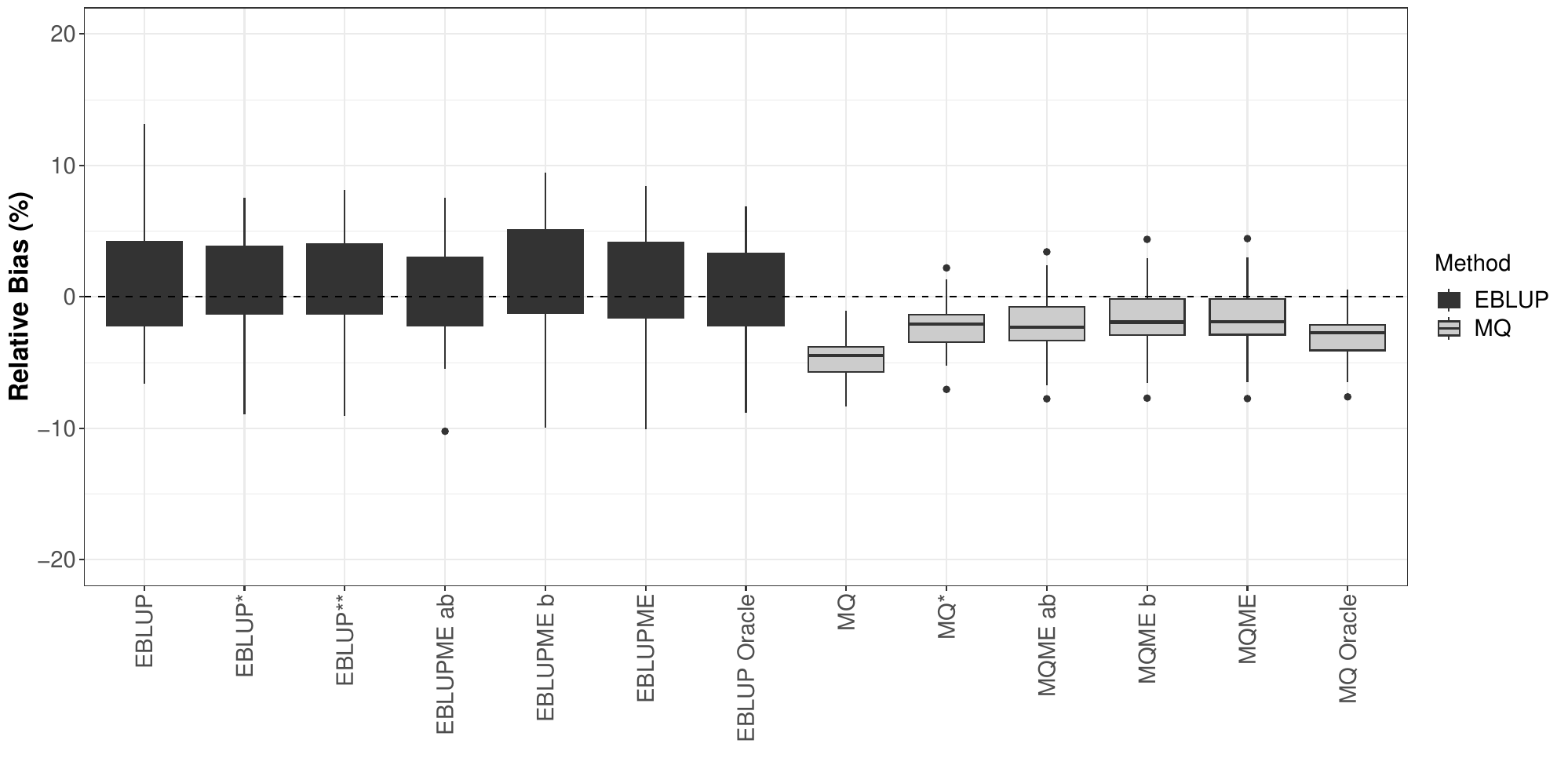}\\
    \includegraphics[scale=0.53]{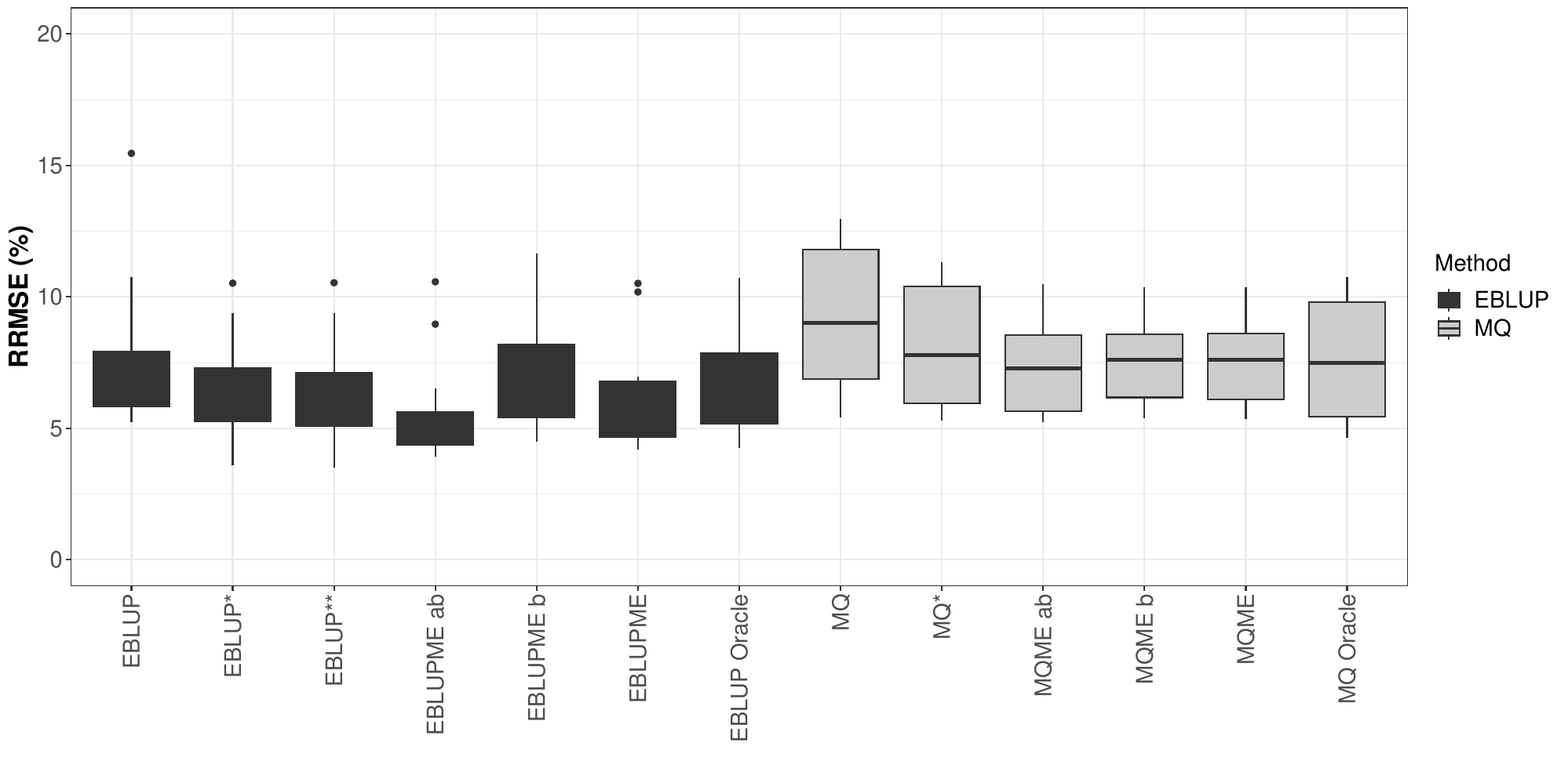}
    \caption{Boxplots showing area-specific values of the RB ($\%$) and the RRMSE ($\%$) of the small area estimators in the design-based experiment. }
    \label{fig:db:rbias:rrmse}
\end{figure}

Figure \ref{fig:db:rbias:rrmse} shows boxplots of area-specific values of the RB (top panel) and the RRMSE (bottom panel) of the small area predictors in the design-based experiment, both reported as percentages. From these plots we can see that the estimators proposed herein ($EBLUPME~ab$, $EBLUPME~b$, $EBLUPME~b$, $MQME~ab$, $MQME~b$ and $MQME$) work well in terms of both bias and RRMSE compared with the unmodified EBLUP and MQ-based predictors that ignore linkage error and the $EBLUP^{\star}$ and $MQ^{\star}$ \citep{salvati2021}. The $EBLUPME~ab$ performs best in terms of bias and efficiency, whereas the $MQME~ab$ is the predictor which performs best in RRMSE within the set of MQ estimators. These results are expected because the $EBLUPME~ab$  and the $MQME~ab$ exploit information that are not used by the other proposed predictors, namely the true mismatch rates and blocks. 
From this perspective, the performance of the estimators $EBLUPME$ and $MQME$ that lack this information is convincing. The reported results also indicate that estimators that correct for linkage error seem to offer the most balanced performance in terms of both bias and MSE for this population.

\begin{figure}[h!]
    \centering
    \includegraphics[scale=0.53]{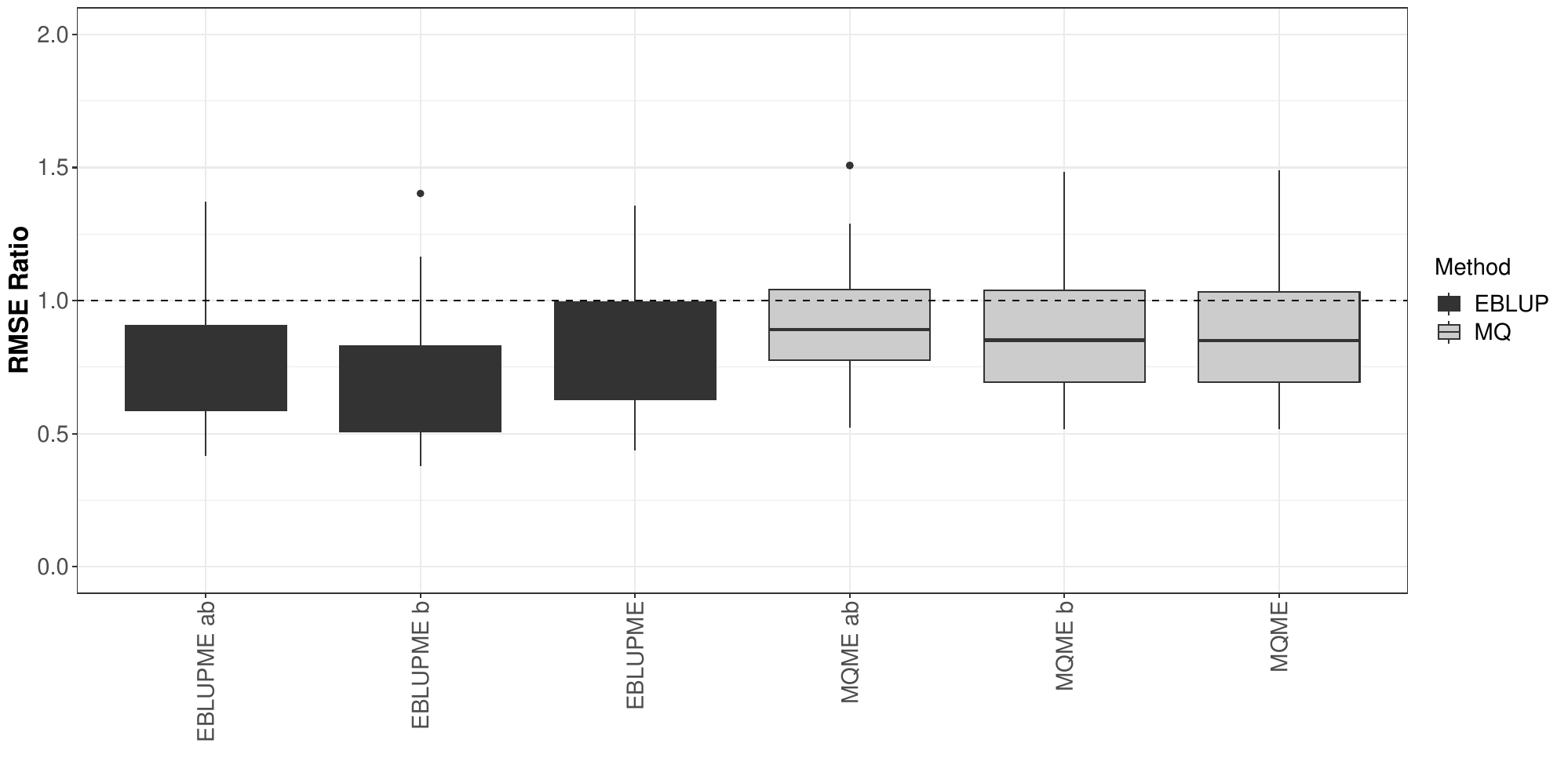}
    \caption{Boxplots showing area-specific values of the RMSE ratios of the MSE estimators in the design-based experiment (the RMSE ratio is defined as the ratio of the average over repeated sampling of the RMSE estimator for an estimator to the actual RMSE of this estimator under repeated sampling.}
    \label{fig:db:rmse}
\end{figure}

Regarding MSE estimation, Figure \ref{fig:db:rmse} shows the area-specific values of the RMSE ratios of the MSE estimators in the design-based experiment. In particular, the RMSE ratio is defined as the ratio of the average estimated RMSE for each area to the actually incurred RMSE. We observe that the MSE estimators proposed in $\S$ \ref{sec:msemq} behave very similarly for MQ-based predictors accounting for linkage error. The resulting estimates align quite well with the actual area-specific MSEs of these predictors. While still roughly capturing the area-specific MSEs of the three associated predictors, the MSE estimates of the EBLUP-type predictor presented in $\S$\ref{sec:MSEEBLUP} exhibit a visible downward bias. Note that the EBLUP-type machinery relies much more strongly on distributional assumptions, which are potentially questionable in this specific application.    

\vspace*{-2.25ex}
\section{Discussion and Conclusions}\label{conclusions}
\vspace*{-1.5ex}
Record linkage is a useful tool for taking advantage of the additional value that can be obtained 
by combining multiple data sources containing complementary pieces of information. Small area estimation is a domain in which record linkage can be utilized to obtain unit-level covariates for creating powerful small area predictors. It is often tacitly assumed that the associated linkage process is free of errors, particularly in the secondary analysis setting studied herein since data analysts may simply rely on the correctness of the linked records provided. In this paper, we have proposed an approach that adjusts two popular regression-based small area predictors for possible mismatch error in the underlying linked data set. The methods we propose can be seen also as a robustification of the two predictors. In fact, false matches can generate non-representative outliers, and we modify the predictors to downweigh them; if outliers are present for other reasons our methods can accommodate them, too.

{\em Missed matches}, i.e., correct links not identified as such, is the second type of error that is presumably equally important. Such missing links are generally not ignorable and not accounting for this possibility can thus introduce selection bias. This prompts the need for methods that can account for both types of errors. Concerning avenues of future research in addition to those pointed out in the body of the paper, we note that the proposed approach relies on strong ignorability assumptions associated with the latent mismatch indicators. It is worth investigating to what extent these assumptions can be relaxed. Second, the discussion in this paper is confined to unit-level small area estimation based on linear regression. Future work could extend our approach to i) outcome variables (e.g., proportions or counts) mandating the use of non-linear regression models, and ii) area-level models such as the popular Fay-Herriot model \citep{Fay1979}.

\clearpage
\bibliographystyle{apalike}
\setlength{\bibsep}{1.6ex}
\bibliography{references_M.bib}

\newpage

\setcounter{equation}{0}
\renewcommand{\theequation}{\thesection.\arabic{equation}}
\appendix

\section{Derivation of the complete data likelihood \eqref{eq:pseudolikelihood}}\label{app:composite_likelihood_derivation}
In this appendix, we provide a step-by-step derivation of the complete data (composite) likelihood \eqref{eq:pseudolikelihood}. We use the symbol $\cong$ for expressing identities justified within the framework of composite likelihood. For example, $f(q_1, q_2) \cong f(q_1) \cdot f(q_2)$, where $f(q_1, q_2)$ is the joint likelihood (PDF) of the quantities $(q_1, q_2)$ whose marginal PDFs are given by $f(q_1)$ and $f(q_2)$, respectively. We have 
\begin{align}
&L(\bm{\beta}, \sigma, \bm{\Sigma}, \bm{\alpha}|\{ \lambda_j, \, \bm{\gamma}_j \}_{j = 1}^D) \notag \\ 
&= f(\{ (y_{ij}^{(s)}, \M{x}_{ij}^{(s)}, \M{z}_{ij}^{(s)}, \M{d}_{ij}^{(s)}, m_{ij}^{(s)}, \bm{\gamma}_j) \}_{i = 1, \, j=1}^{n_j, \;\, D}) \notag \\
&\overset{\circled{1}}{\cong} \prod_{j = 1}^D f(\{ (y_{ij}^{(s)}, \M{x}_{ij}^{(s)}, \M{z}_{ij}^{(s)}, \M{d}_{ij}^{(s)}, m_{ij}^{(s)}, \bm{\gamma}_j) \}_{i = 1}^{n_j}) \notag \\
&\overset{\circled{2}}{\cong} \prod_{j = 1}^D \Big[ f(\{ (y_{ij}^{(s)}, \M{x}_{ij}^{(s)}, \M{z}_{ij}^{(s)}, \M{d}_{ij}^{(s)}, m_{ij}^{(s)}, \bm{\gamma}_j) \}_{i \in \lambda_j}) \times f(\{ (y_{ij}^{(s)}, \M{x}_{ij}^{(s)}, \M{z}_{ij}^{(s)}, \M{d}_{ij}^{(s)}, m_{ij}^{(s)}, \bm{\gamma}_j) \}_{i \in \overline{\lambda}_j}) \Big] \notag \\
&\overset{\circled{3}}{=} \prod_{j = 1}^D  \Big[ f \Big(\{ y_{ij}^{(s)}  \}_{i \in \lambda_j} \Big| \{ (\M{x}_{ij}^{(s)}, \M{z}_{ij}^{(s)}, m_{ij}^{(s)}, \M{d}_{ij}^{(s)}, \bm{\gamma}_j) \}_{i \in \lambda_j} \Big) \times \notag \\
&\quad \qquad \times f\Big(\{ y_{ij}^{(s)} \}_{i \in \overline{\lambda}_j} \Big | \{ (\M{x}_{ij}^{(s)}, \M{z}_{ij}^{(s)}, m_{ij}^{(s)}, \M{d}_{ij}^{(s)}, \bm{\gamma}_j) \}_{i \in \overline{\lambda}_j} \Big) \times f( \{ (\M{x}_{ij}^{(s)}, \M{z}_{ij}^{(s)}, m_{ij}^{(s)}, \M{d}_{ij}^{(s)} \}) \times f(\bm{\gamma}_j) \Big ] \notag \notag \\
&\overset{\circled{4}}{\propto}  \prod_{j = 1}^D  \Big[ f \Big(\{ y_{ij}^{(s)}  \}_{i \in \lambda_j} \Big| \{ (\M{x}_{ij}^{(s)}, \M{z}_{ij}^{(s)}, \bm{\gamma}_j) \}_{i \in \lambda_j} \Big)  \times f \Big(\{ y_{ij}^{(s)}  \}_{i \in \overline{\lambda}_j} \Big) \times \notag \\
&\quad \qquad \times f(\{ m_{ij}^{(s)} \}_{i \in \lambda_j} | \{ \M{d}_{ij}^{(s)} \}_{i \in \lambda_j})
\times f(\{ m_{ij}^{(s)} \}_{i \in \overline{\lambda}_j} | \{ \M{d}_{ij}^{(s)} \}_{i \in \overline{\lambda}_j}) \times f(\bm{\gamma}_j) \Big]
\notag \\
&\overset{\circled{5}}{\cong} \prod_{j = 1}^D \bigg\{  (2 \pi \sigma^2)^{-|\lambda_j|/2} \exp \left(-\frac{\nnorm{\M{y}_{\lambda_j}^{(s)} - \M{X}_{\lambda_j}^{(s)} \bm{\beta} - \M{Z}_{\lambda_j}^{(s)} \bm{\gamma}_j}_2^2}{2 \sigma^2} \right) \times \notag \\
   & \quad \qquad \times \left[ |\bm{\Sigma}|^{-1/2} (2 \pi)^{q/2} \exp \left(-\frac{1}{2} \bm{\gamma}_j^{\T} \bm{\Sigma}^{-1} \bm{\gamma}_j \right) \right] \times \prod_{i \in \overline{\lambda}_j} g(y_{ij}^{(s)}) \times \notag \\
   &\quad \qquad \times \prod_{i \in \lambda_j} \p(m_{ij}^{(s)} = 0 | \M{d}_{ij}^{(s)}; \bm{\alpha})  \times \prod_{i \in \overline{\lambda}_j} \p(m_{ij}^{(s)} = 1 | \M{d}_{ij}^{(s)}; \bm{\alpha}) \bigg\} \label{eq:pseudolikelihood_app},
\end{align}
Step $\circled{1}$ produces a factorization over the $D$ areas, with the symbol $\cong$ used here for the reasons explained above since mismatch error may occur across areas. In step $\circled{2}$, terms are further divided according to match status. Step $\circled{3}$ is obtained by conditioning and then using independence of 
the random effects $\{ \bm{\gamma}_j \}_{j = 1}^D$ from the quantities $\{ (\M{x}_{ij}^{(s)}, \M{z}_{ij}^{(s)}, m_{ij}^{(s)}, \M{d}_{ij}^{(s)}) \}$. In step $\circled{4}$ we invoke i) the independence model for mismatches (cf.~last paragraph of $\S$\ref{sec:meth}), ii) the assumption that linkage error is independent of the outcomes and the covariates $\M{x}$ and $\M{z}$, and iii) the unit-level SA model \eqref{eq:unitlevel_SA} for correct matches, which implies that for those observations $\{ \M{d}_{ij}^{(s)} \}$ after the conditioning symbol $|$ can be dropped. Finally, in step $\circled{5}$ the first line and the term inside square brackets in the second line result from substituting the specific form of the likelihood for correct matches and the random effects $\{ \bm{\gamma}_j \}_{j = 1}^D$, respectively, according to \eqref{eq:unitlevel_SA}. The terms in the third factor $\prod_{i \in \overline{\lambda}_j} g(y_{ij}^{(s)})$ represent the individual likelihood contributions of mismatched observations whose PDF is given by \eqref{eq:marginal_ultimate}. Specifically, under the linear mixed effect model \eqref{eq:unitlevel_SA}, the PDF $g$ has the following representation:
\begin{equation*}
g(y) = \sum_{j = 1}^D \sum_{i = 1}^{N_j} \; \frac{\p(\texttt{s}_{ij} | \M{x}_{ij}, \M{z}_{ij})}{\sum_{\ell = 1}^{D} \sum_{k = 1}^{N_{\ell}} \p(\texttt{s}_{k\ell} | \M{x}_{k\ell}, \M{z}_{k\ell})} \, 
\frac{1}{\sigma_{ij}^{*}} \, \varphi\left(\frac{y  \, - \mu_{ij}^{*}}{\sigma_{ij}^{*}} \right), \; \; y \in \R,
\end{equation*}
where $\mu_{ij}^* \coloneq \M{x}_{ij}^{\T} \bm{\beta}_*$ and 
$\sigma_{ij}^* \coloneq (\M{z}_{ij}^{\T} \bm{\Sigma}_* \M{z}_{ij} + \sigma_*^2)^{1/2}$, $1 \leq i \leq N_j$, $1 \leq j \leq D$, with 
$\varphi$ denoting the PDF of the $N(0,1)$-distribution.

The final line in \eqref{eq:pseudolikelihood_app} represents the individual likelihood contributions for the mismatch indicators. Factorization over $\lambda_j$, $\overline{\lambda}_j$, and over $j = 1,\ldots,D$ in \eqref{eq:pseudolikelihood_app} is justified in terms of composite likelihood: while products over the terms 
$\{ g(y_{ij}^{(s)}) \}$, $\{ \p(m_{ij}^{(s)} = 0 | \M{d}_{ij}^{(s)}; \bm{\alpha}) \}$ and 
$\{ \p(m_{ij}^{(s)} = 1 | \M{d}_{ij}^{(s)}; \bm{\alpha}) \}$ may not give rise to {\em joint} likelihood contributions, each factor represents a valid marginal (i.e., unit-level) contribution.

\section{Calculations for the EM algorithm for the approach based on the SA linear mixed effect model}\label{app:EMalg}

\subsection{Exact EM Algorithm}\label{subsec:EM_exact}
\noindent{\bfseries E-step}. In the following, we evaluate the expectation of the complete data (pseudo-) log likelihood according to \eqref{eq:pseudolikelihood} where the expectation is conditional on the data $\mc{D} = \{ (y_{ij}^{(s)}, \M{x}_{ij}^{(s)}, \M{z}_{ij}^{(s)}), \M{d}_{ij}^{(s)} \}$ and the parameters fixed as $\bm{\theta}^{(t)} = \{ \bm{\beta}^{(t)}, \sigma^{{2} \, (t)},  \bm{\Sigma}^{(t)}, \bm{\alpha}^{(t)} \}$, where the superscript $^{(t)}$ here refers to an iteration counter. To avoid notational clutter, we henceforth drop the superscript $^{(s)}$ whenever the reference to sampled units is clear from the context. For a given function $\phi$, the E-step will require the evaluation of expectations of the form 
\begin{align}
\E_{\lambda_j, \bm{\gamma}_j}[\phi(\lambda_j, \bm{\gamma}_j)|\mc{D};\bm{\theta}^{(t)}] &= \int_{\R^q} \sum_{\ell_j \in \mc{P}_j} \phi(\ell_j, \bm{\gamma}_j) \, f(\bm{\gamma}_j|\mc{D}, \{ \lambda_j = \ell_j \}; \bm{\theta}^{(t)}) \, \p(\lambda_j = \ell_j | \mc{D}; \bm{\theta}^{(t)}) \;d\bm{\gamma}_j \notag \\[1ex]
  &= \sum_{\ell_j \in \mc{P}_j} \left\{ \p(\lambda_j = \ell_j | \mc{D}; \bm{\theta}^{(t)}) \int_{\R^q} \phi(\ell_j, \bm{\gamma}_j) \,  f(\bm{\gamma}_j|\mc{D}, \{ \lambda_j = \ell_j \}; \bm{\theta}^{(t)}) \; d\bm{\gamma}_j \right\}, \notag \\
 &\quad \; 1 \leq j \leq D. \label{eq:Esplit_basic_decomposition}
\end{align}
We start by computing
\begin{align}\label{eq:Pell_1}
  \p(\lambda_j = \ell_j | \mc{D}; \bm{\theta}^{(t)}) = \frac{f(\mc{D} | \lambda_j = \ell_j; \bm{\theta}^{(t)}) \p(\lambda_j = \ell_j; \bm{\theta}^{(t)})}{\sum_{\wt{\ell}_j \in \mc{P}_j} f(\mc{D} | \lambda_j = \wt{\ell}_j; \bm{\theta}^{(t)}) \p(\lambda_j = \wt{\ell}_j; \bm{\theta}^{(t)})}, \quad \ell_j \in \mc{P}_j, \; 1 \leq j \leq D,
 \end{align}
where 
\begin{align}\label{eq:Pell_2}
\begin{split}
&f(\mc{D} | \lambda_j = \ell_j; \bm{\theta}^{(t)}) \p(\lambda_j = \ell_j; \bm{\theta}^{(t)}) \\
&\cong (2\pi)^{-|\ell_j|/2} |\bm{\Lambda}_{\ell_j}^{(t)}|^{-1/2} \exp\left(-\frac{1}{2} (\M{y}_{\ell_j} - \M{X}_{\ell_j} \bm{\beta}^{(t)})^{\T} [\bm{\Lambda}_{\ell_j}^{(t)}]^{-1} (\M{y}_{\ell_j} - \M{X}_{\ell_j} \bm{\beta}^{(t)}) \right) \times   \\
&\quad \times \prod_{i \in \ell_j} (1 - h(\M{d}_{ij};\bm{\alpha}^{(t)})) \times \prod_{i \in \overline{\ell}_j} [ g(y_{ij}) \times h(\M{d}_{ij};\bm{\alpha}^{(t)})], \quad \ell_j \in \mc{P}_j, \; 1 \leq j \leq D,
\end{split}
\end{align}
where the use of $\cong$ is addressed in a remark at the end of this subsection. In the above equation, we have used the following additional symbols: 
\begin{itemize}
\item $\overline{\ell}_j = \{1, \ldots, n_j \} \setminus \ell_j$, i.e., the complement of $\ell_j \in \mc{P}_j$,
\item $\bm{\Lambda}_{\ell_j}^{(t)} = \M{Z}_{\ell_j} \bm{\Sigma}^{(t)} \M{Z}_{\ell_j}^{\T} + \sigma^{2 \, (t)} I_{|\ell_j|}$ denotes the covariance matrix of $\M{y}_{\ell_j}$ conditional on $\{ \lambda_j = \ell_j \}$, $1 \leq j \leq D$.    
\end{itemize}
Note that as before, superscripts $\ell_j$ in $\M{y}$, $\M{X}$ and $\M{Z}$ refer to the corresponding subvectors and submatrices, respectively. 

As a last ingredient, we use that according to standard rules $\bm{\gamma}_j | \mc{D}, \{ \lambda_j = \ell_j \};\bm{\theta}^{(t)}$ follows a Normal distribution with 
mean $\bm{\mu}_{\ell_j}^{(t)}$ and covariance $\bm{\Sigma}_{\ell_j}^{(t)}$ given by  
\begin{align}\label{eq:conddist_gamma}
\begin{split}
&\bm{\gamma}_j | \mc{D}, \{ \lambda_j = \ell_j  \} \sim N\left(\bm{\mu}_{\ell_j}^{(t)}, \bm{\Sigma}_{\ell_j}^{(t)} \right)  \\
&\bm{\mu}_{\ell_j}^{(t)} \coloneq \bm{\Sigma}^{(t)} \M{Z}_{\ell_j}^{\T} \left( \M{Z}_{\ell_j} \bm{\Sigma}^{(t)} \M{Z}_{\ell_j}^{\T} + \sigma^{2\,(t)}  I_{|\ell_j|}\right)^{-1} (\M{y}_{\ell_j} - \M{X}_{\ell_j} \bm{\beta}^{(t)}), \\
&\bm{\Sigma}_{\ell_j}^{(t)} \coloneq  \bm{\Sigma}^{(t)} - \bm{\Sigma}^{(t)} \M{Z}_{\ell_j}^{\T} \left( \M{Z}_{\ell_j} \bm{\Sigma}^{(t)} \M{Z}_{\ell_j}^{\T} + \sigma^{2\,(t)} I_{|\ell_j|}\right)^{-1} \M{Z}_{\ell_j} \bm{\Sigma}^{(t)}.  
\end{split}
\end{align}
Equipped with \eqref{eq:Esplit_basic_decomposition}, \eqref{eq:Pell_1}, \eqref{eq:Pell_2} and \eqref{eq:conddist_gamma}, we are 
in position to evaluate the expected complete data negative (pseudo)-log-likelihood corresponding to \eqref{eq:integrated_likelihood}. Taking the negative logarithm of the complete data (pseudo)-likelihood \eqref{eq:pseudolikelihood}, we obtain the following expression after dropping additive constants not depending on the parameters $\bm{\theta}$: 
\begin{align}\label{eq:completedata_negloglik}
\sum_{j = 1}^D  &\Bigg[ \-\frac{|\lambda_j|}{2} \log(\sigma^2) + \frac{\nnorm{\M{y}_{\lambda_j} - \M{X}_{\lambda_j} \bm{\beta}}_2^2}{2 \sigma^2} + \frac{\nnorm{\M{Z}_{\lambda_j} \bm{\gamma}_j}_2^2}{2 \sigma^2} - \frac{\scp{\M{y}_{\lambda_j} -
    \M{X}_{\lambda_j} \bm{\beta}}{\M{Z}_{\lambda_j} \bm{\gamma}_j}}{\sigma^2}  +\frac{1}{2} \log |\bm{\Sigma}|  +\frac{1}{2} \bm{\gamma}_j^{\T} \bm{\Sigma}^{-1} \bm{\gamma}_j  + \notag \\
  &- \sum_{i \in \lambda_j}  \log\left(1 - h(\M{d}_{ij};\bm{\alpha}) \right) - \sum_{i \in \overline{\lambda}_j} \log\left(h(\M{d}_{ij};\bm{\alpha}) \right)  \Bigg] 
\end{align}
Evaluating the expectation of that expression according to \eqref{eq:Esplit_basic_decomposition}, we obtain that
\begin{align}
  &\sum_{j = 1}^N \sum_{\ell_j \in \mc{P}_j} \p(\lambda_j = \ell_j | \mc{D}; \bm{\theta}^{(t)}) \Bigg [ \frac{|\ell_j|}{2} \log(\sigma^2) +\frac{\nnorm{\M{y}_{\ell_j} - \M{X}_{\ell_j} \bm{\beta}}_2^2}{2 \sigma^2} +\frac{\E[\nnorm{\M{Z}_{\ell_j} \bm{\gamma}_j}_2^2 | \mc{D}, \{\lambda_j = \ell_j \}; \bm{\theta}^{(t)}]}{2 \sigma^2} - \notag \\
  & \qquad \qquad\qquad \qquad \qquad \qquad -\frac{\scp{\M{y}_{\ell_j} -
    \M{X}_{\ell_j} \bm{\beta}}{\M{Z}_{\ell_j} \E[\bm{\gamma}_j |  \mc{D}, \{\lambda_j = \ell_j \}; \bm{\theta}^{(t)}]}}{\sigma^2} \, +  \notag\\
  &  \qquad \qquad\qquad \qquad \qquad \qquad  +\frac{1}{2} \log |\bm{\Sigma}|  +\frac{1}{2} \E[\bm{\gamma}_j^{\T} \bm{\Sigma}^{-1} \bm{\gamma}_j|  \mc{D}, \{\lambda_j = \ell_j \}; \bm{\theta}^{(t)}]  \, - \notag\\
  &\qquad \qquad \qquad \qquad \qquad \qquad - \sum_{i \in \ell_j}  \log\left(1 - h(\M{d}_{ij};\bm{\alpha}) \right) - \sum_{i \in \overline{\ell}_j} \log\left(h(\M{d}_{ij};\bm{\alpha}) \right) \Bigg]. \label{eq:expected_negloglik}
\end{align}
Using \eqref{eq:conddist_gamma}, the remaining expectations inside the sum can be evaluated as follows: 
\begin{align*}
&\E[\bm{\gamma}_j |  \mc{D}, \{\lambda_j = \ell_j \}; \bm{\theta}^{(t)}] = \bm{\mu}_{\ell_j}^{(t)}, \\
&\E[\bm{\gamma}_j^{\T} \bm{\Sigma}^{-1} \bm{\gamma}_j|  \mc{D}, \{\lambda_j = \ell_j \}; \bm{\theta}^{(t)}] = \tr(\bm{\Sigma}^{-1} \E[\bm{\gamma}_j \bm{\gamma}_j^{\T} |  \mc{D}, \{\lambda_j = \ell_j \}; \bm{\theta}^{(t)}]),\\
&\E[\nnorm{\M{Z}_{\ell_j} \bm{\gamma}_j}_2^2 | \mc{D}, \{\lambda_j = \ell_j \}; \bm{\theta}^{(t)}] = \tr(\M{Z}_{\ell_j}^{\T} \M{Z}_{\ell_j} \E[\bm{\gamma}_j \bm{\gamma}_j^{\T} | \mc{D}, \{\lambda_j = \ell_j \}; \bm{\theta}^{(t)}]) \\
&\qquad \qquad \qquad \qquad \qquad \qquad \;\;\,=\sum_{i \in \ell_j} \M{z}_{ij}^{\T} \E[\bm{\gamma}_j \bm{\gamma}_j^{^{\T}} | \mc{D}, \{\lambda_j = \ell_j \}; \bm{\theta}^{(t)}]) \M{z}_{ij} ,    
\end{align*}
where 
\begin{equation}\label{eq:Egammagammat}
\E[\bm{\gamma}_j\bm{\gamma}_j^{\T}  | \mc{D}, \{\lambda_j = \ell_j \}; \bm{\theta}^{(t)}] = \bm{\Sigma}_{\ell_j}^{(t)} + \bm{\mu}_{\ell_j}^{(t)} \left[ \bm{\mu}_{\ell_j}^{(t)} \right]^{\T}. 
\end{equation}
This concludes the derivation of the E-step. The resulting M-step is explained below. 
\vskip1ex
\noindent{\bfseries M-step}. The goal is to minimize the expected negative complete data (pseudo-) log-likelihood \eqref{eq:expected_negloglik}. It can be seen that the updates for $\bm{\beta}$, $\sigma^2$, and $\bm{\Sigma}$ can be expressed in closed form. Specifically, we obtain 
\begin{align*}
&\wh{\bm{\beta}}^{(t+1)} = \argmin_{\bm{\beta}} \left\{ \sum_{j = 1}^D \sum_{\ell_j \in \mc{P}_j \setminus \emptyset} \p(\lambda_j = \ell_j | \mc{D}; \bm{\theta}^{(t)}) \, \nnorm{\M{y}_{\ell_j} - \M{Z}_{\ell_j} \bm{\mu}_{\ell_j}^{(t)} - \M{X}_{\ell_j} \bm{\beta}}_2^2 \right\},  \\[1ex]
&\mbox{{\footnotesize$\sigma^{2\,(t+1)} = \displaystyle\frac{\sum_{j = 1}^D \sum_{\ell_j \in \mc{P}_j \setminus \emptyset} \p(\lambda_j = \ell_j | \mc{D};\bm{\theta}^{(t)}) \left\{ \nnorm{\M{y}_{\ell_j} - \M{X}_{\ell_j} \wh{\bm{\beta}}^{(t+1)}}_2^2 + \sum_{i \in \ell_j} \M{z}_{ij}^{\T} \E[\bm{\gamma}_j \bm{\gamma}_{j}^{\T} | \mc{D}, \{\ell_j = \lambda_j\}; \bm{\theta}^{(t)}] \M{z}_{ij} - 2 C_{\ell_j}^{(t)} \right\}}{\sum_{j = 1}^D \sum_{\ell_j \in \mc{P}_j \setminus \emptyset} |\ell_j| \p(\lambda_j = \ell_j | \mc{D};\bm{\theta}^{(t)})}$}}, \\[.5ex] 
&\qquad \mbox{{\footnotesize $C_{\ell_j}^{(t)} \coloneq \nscp{\M{y}_{\ell_j} - \M{X}_{\ell_j} \wh{\bm{\beta}}^{(t+1)}}{\M{Z}_{\ell_j} \bm{\mu}_{\ell_j}^{(t)}}$}} \\[2ex]
&\bm{\Sigma}^{(t+1)} =  \frac{\sum_{j = 1}^D \sum_{\ell_j \in \mc{P}_j} \p(\lambda_j = \ell_j | \mc{D}; \bm{\theta}^{(t)}) \E[\bm{\gamma}_j \bm{\gamma}_j^{\T} |  \mc{D}, \{\lambda_j = \ell_j \}; \bm{\theta}^{(t)}]}{\sum_{j = 1}^D \sum_{\ell_j \in \mc{P}_j} \p(\lambda_j = \ell_j | \mc{D}; \bm{\theta}^{(t)})}
\end{align*}
As stated, the update of $\bm{\beta}$ involves $\sum_{k = 1}^{n_j} k \binom{n_j}{k} = n_j 2^{n_j - 1}$ terms for each $j=1,\ldots,D$. However, a reduction to $n_j$ terms can be achieved as follows. Expanding the square and dropping terms not depending on $\bm{\beta}$, we obtain 
\begin{align*}
&\sum_{j = 1}^D \sum_{\ell_j \in \mc{P}_j \setminus \emptyset} \p(\lambda_j = \ell_j | \mc{D}; \bm{\theta}^{(t)}) \, \left\{ \bm{\beta}^{\T} \M{X}_{\ell_j}^{\T}  \M{X}_{\ell_j} \bm{\beta} - 2 \bm{\beta}^{\T} \M{X}_{\ell_j}^{\T} (\M{y}_{\ell_j} - \M{Z}_{\ell_j} \bm{\mu}_{\ell_j}^{(t)}) \right\} \\
&= \sum_{j = 1}^D \sum_{\ell_j \in \mc{P}_j \setminus \emptyset} \, \p(\lambda_j = \ell_j | \mc{D}; \bm{\theta}^{(t)}) \, \sum_{i \in \ell_j} \{ \bm{\beta}^{\T} \M{x}_{ij} \M{x}_{ij}^{\T} \bm{\beta} - 2 \bm{\beta}^{\T} \M{x}_{ij} (y_{ij} - \M{z}_{ij}^{\T} \bm{\mu}_{\ell_j}^{(t)}) \} \\
&= \sum_{j = 1}^D \sum_{i = 1}^{n_j} \sum_{\{ \ell_j \in \mc{P}_j: \; i \in \ell_j \}} \p(\lambda_j = \ell_j | \mc{D}; \bm{\theta}^{(t)}) \{ \bm{\beta}^{\T} \M{x}_{ij} \M{x}_{ij}^{\T} \bm{\beta} - 2 \bm{\beta}^{\T} \M{x}_{ij} (y_{ij} - \M{z}_{ij}^{\T} \bm{\mu}_{\ell_j}^{(t)})  \} \\
&= \sum_{j = 1}^D \sum_{i = 1}^{n_j} \left\{ \bm{\beta}^{\T} \M{x}_{ij} \omega_{ij}^{(t)} \M{x}_{ij}^{\T} \bm{\beta} -2\bm{\beta}^{\T} \M{x}_{ij} \omega_{ij}^{(t)} y_{ij} + 2\bm{\beta}^{\T} \M{x}_{ij} \M{z}_{ij}^{\T} \overline{\bm{\mu}}_{ij}^{(t)} \right\}, \\
&\quad \omega_{ij}^{(t)} \coloneq  \sum_{\{ \ell_j \in \mc{P}_j: \; i \in \ell_j \}}
\p(\lambda_j = \ell_j | \mc{D}; \bm{\theta}^{(t)}), \quad \overline{\bm{\mu}}_{ij}^{(t)} \coloneq \sum_{\{ \ell_j \in \mc{P}_j: \; i \in \ell_j \} }  \p(\lambda_j = \ell_j | \mc{D}; \bm{\theta}^{(t)}) \, \bm{\mu}_{\ell_j}^{(t)}. 
\end{align*}
Converting the last expression back to a matrix-vector representation, the update for $\bm{\beta}$ can be expressed as the solution of the following linear system of equations: 
\begin{equation}\label{eq:update_beta_rewritten}
(\M{X}^{\T} \M{W}^{(t)} \M{X}) \wh{\bm{\beta}}^{(t+1)} = \M{X}^{\T} \M{W}^{(t)} \M{y} - \M{X}^{\T} \bm{\xi}^{(t)}, \quad \bm{\xi}^{(t)} = (\M{z}_{ij}^{\T} \overline{\bm{\mu}}_{ij}^{(t)}),
\end{equation}
where $\M{W}^{(t)}$ is a diagonal matrix with diagonal entries $\{ \omega_{ij}^{(t)} \}$.  
The update for $\sigma^2$ can be re-expressed similarly. Note that the update for $\bm{\Sigma}$ is simply a weighted
average of the terms \eqref{eq:Egammagammat}. 

Finally, observe that according to the structure \eqref{eq:expected_negloglik}, $\bm{\alpha}$ can be updated separately from the other parameters. We have 
\begin{align}
\wh{\bm{\alpha}}^{(t+1)} &= \argmin_{\bm{\alpha}} \left\{ \sum_{j = 1}^D \sum_{\ell_j \in \mc{P}_j} \p(\lambda_j = \ell_j | \mc{D}; \bm{\theta}^{(t)}) \left[- \sum_{i \in \ell_j}  \log\left(1 - h(\M{d}_{ij};\bm{\alpha}) \right) - \sum_{i \in \overline{\ell}_j} \log\left(h(\M{d}_{ij};\bm{\alpha}) \right) \right]\right \} \notag \\
                         &= \argmin_{\bm{\alpha}} \left\{ -\sum_{j = 1}^D \sum_{i = 1}^{n_j} \left[ \omega_{ij}^{(t)}  \log\left(1 - h(\M{d}_{ij};\bm{\alpha}) \right) + (1 - \omega_{ij}^{(t)}) \log\left(h(\M{d}_{ij};\bm{\alpha}) \right) \right] \right \} \label{eq:update_alpha_rewritten}
\end{align}
where the identity of the two expressions follows along the same lines as the above re-formulation of the update for 
$\bm{\beta}$. As an example, consider the perhaps simplest case in which $h \equiv \alpha \in (0,1)$ is a constant. In this case,
the update for $\alpha$ is seen to be the average of the $\{ 1 - \omega_{ij}^{(t)} \}$.
\vskip1ex
\noindent {\bfseries Remark}. We note that the EM algorithm is a simply a computational vehicle (a specific majorization--minimization (MM) algorithm, cf.~\cite{Hunter2004}) for optimizing likelihood functions involving integrals over latent variables. As such, it is immaterial whether the function to be optimized arises from a ``genuine" likelihood
or from a composite likelihood, and evaluation of $\p(\lambda_j = \ell_j;\bm{\theta}^{(t)})$, $1 \leq j \leq D$, in the E-step according to \eqref{eq:Pell_2} is justified.  

\subsection{Monte Carlo Approximation (MC-EM)}\label{subsec:MCEM}
In this subsection, we provide details on the approximation of integrals \eqref{eq:Esplit_basic_decomposition} in the E-step via Monte Carlo integration in the spirit of \cite{Wei1990}, i.e., sampling from the distribution of the latent variables $\{ \bm{\gamma}_j, \lambda_j \}_{j = 1}^D$ given 
$\mc{D}$ and the current parameters $\bm{\theta}^{(t)}$. As a result, we obtain an approximation to the expected complete data negative (pseudo)-log-likelihood \eqref{eq:expected_negloglik}, side-stepping the combinatorial barrier that is associated with an exact E-step computation as discussed in the previous subsection. Sampling from the joint distribution of 
$\bm{\gamma}_j, \lambda_j | \mc{D}; \bm{\theta}^{(t)}$ is facilitated via Gibbs sampling: we alternate between sampling 
of (i) $\lambda_j | \bm{\gamma}_j, \mc{D}; \bm{\theta}^{(t)}$ and (ii) $\bm{\gamma}_j | \lambda_j, \mc{D}; \bm{\theta}^{(t)} $, $1 \leq j \leq D$. In fact, as we elaborate below, both (i) and (ii) reduce to straightforward tasks: the former reduces to independent sampling  of mismatch indicators $\{ m_{ij} \}$, whereas the latter involves sampling from a normal distribution. In particular, the ability to reduce sampling of $\{ \lambda_j \}$ to independent (Bernoulli) sampling of mismatch indicators breaks the combinatorial barrier that is associated with the exact E-step. 
\vskip1ex
{\em Sampling from $\lambda_j | \bm{\gamma}_j, \mc{D}; \bm{\theta}^{(t)}$}. We have 
\begin{align}
f(\lambda_j | \mc{D}, \bm{\gamma}_j;\bm{\theta}^{(t)}) &= f(\{ m_{ij} \}_{i = 1}^{n_j} | \mc{D}, \bm{\gamma}_j;\bm{\theta}^{(t)}) \notag \\ &\propto f(\mc{D}|\bm{\gamma}_j, \{ m_{ij} \}_{i = 1}^{n_j};\bm{\theta}^{(t)}) \cdot f(\{ m_{ij} \}_{i = 1}^{n_j} | \bm{\gamma}_j; \bm{\theta}^{(t)}) \notag\\
&= f(\mc{D}|\bm{\gamma}_j, \{ m_{ij} \}_{i = 1}^{n_j};\bm{\theta}^{(t)}) \cdot f(\{ m_{ij} \}_{i = 1}^{n_j} | \bm{\theta}^{(t)}) \notag\\
&\overset{\circled{1}}{=}\prod_{i = 1}^{n_j} f(y_{ij} | \M{x}_{ij}, \M{z}_{ij}, \bm{\gamma}_j,  m_{ij}; \bm{\theta}^{(t)}) \times  f( \{ m_{ij} \}_{i = 1}^{n_j} | \bm{\theta}^{(t)}) \notag \\
&\overset{\circled{2}}{\cong} \prod_{i = 1}^{n_j} \left\{ f(y_{ij} | \M{x}_{ij}, \M{z}_{ij}, \bm{\gamma}_j,  m_{ij}; \bm{\theta}^{(t)}) \times  f( m_{ij} | \bm{\theta}^{(t)}) \right\} \label{eq:gibbs_lambda}
\end{align}
where in $\circled{1}$ we have used that the $\{ y_{ij} \}$ are independent given $\bm{\gamma}_j$ and $\circled{2}$ is justified by the remark at the end of $\S$\ref{subsec:EM_exact}. Each term inside the curly bracket in \eqref{eq:gibbs_lambda} is of the form \begin{equation*}
\begin{cases}\frac{1}{\sqrt{2 \pi \sigma^2}}\exp \left(-\frac{(y_{ij} - \M{x}_{ij}^{\T} \bm{\beta}^{(t)} - \M{z}_{ij}^{\T} \bm{\gamma}_j)^2}{2 \sigma^{2(t)}} \right)  \times \left(1 - h(\M{d}_{ij};\bm{\alpha}^{(t)}) \right) \quad &\text{if} \; m_{ij} = 0, \\
g(y_{ij}) \times  h(\M{d}_{ij};\bm{\alpha}^{(t)}) \quad &\text{if} \; m_{ij} = 1.
\end{cases}
\end{equation*}
From the above derivation, it follows that the $\{m_{ij} \}_{i = 1}^{n_j} | \mc{D}, \bm{\gamma}_j;\bm{\theta}^{(t)}$ can be sampled as 
Bernoulli random variables whose probabilities of success are given by 
\begin{equation*}
\frac{g(y_{ij}) \times  h(\M{d}_{ij};\bm{\alpha}^{(t)})}{g(y_{ij}) \times  h(\M{d}_{ij};\bm{\alpha}^{(t)}) + \frac{1}{\sqrt{2 \pi \sigma^2}}\exp \left(-\frac{(y_{ij} - \M{x}_{ij}^{\T} \bm{\beta}^{(t)} - \M{z}_{ij}^{\T} \bm{\gamma}_j)^2}{2 \sigma^{2(t)}} \right)  \times \left(1 - h(\M{d}_{ij};\bm{\alpha}^{(t)}) \right) }, \quad 1 \leq i \leq n_j.     
\end{equation*}
\vskip1ex
{\em Sampling from $\bm{\gamma}_j|\lambda_j, \mc{D}; \bm{\theta}^{(t)}$}. In light of the discussion in $\S$\ref{subsec:EM_exact}, this part does not require further explanation. The distribution of interest is given by a Normal distribution as in \eqref{eq:conddist_gamma}. 
\vskip1ex
{\em M-step}. After a ``burn-in" period the above Gibbs sampling scheme produces samples\\ $\{ \, \{ (m_{ij}^{[u]})_{i = 1}^{n_j}, \bm{\gamma}_j^{[u]} \}_{j = 1}^D \, \}_{u = 1}^U$, where $U$ denotes the total number of samples. In the following M-step, expectations are replaced by their corresponding empirical averages. Specifically, we obtain the following updates: 
\begin{align*}
&(\M{X}^{\T} \wt{\M{W}}^{(t)} \M{X}) \wh{\bm{\beta}}^{(t+1)} = \M{X}^{\T} \wt{\M{W}}^{(t)} \M{y} - \M{X}^{\T} \wt{\bm{\xi}}^{(t)}, \quad \wt{\bm{\xi}}^{(t)} = \left( \sum_{u = 1}^U (1 - m_{ij}^{[u]}) \M{z}_{ij}^{\T} \bm{\gamma}_j^{[u]} \right), \\
& \sigma^{2\,(t + 1)} = \frac{\sum_{j = 1}^D \sum_{i = 1}^{n_j} \left\{ \wt{\omega}_{ij}^{(t)} (y_{ij} - \M{x}_{ij}^{\T} \wh{\bm\beta}^{(t+1)})^2  +  \frac{1}{U} \sum_{u = 1}^U (1 - m_{ij}^{[u]}) \left( \M{z}_{ij}^{\T} \bm{\gamma}_j^{[u]} \right)^2 - 2 (y_{ij} - \M{x}_{ij}^{\T} \wh{\bm \beta}^{(t + 1)})\wt{\xi}_{ij}^{(t)} \right\}}{\sum_{j = 1}^D \sum_{i = 1}^{n_j} \wt{\omega}_{ij}^{(t)}} \\
& \bm{\Sigma}^{(t+1)} = \frac{1}{U \cdot D}\sum_{u = 1}^U \sum_{j= 1}^D \bm{\gamma}_j^{[u]} \big[ \bm{\gamma}_j^{[u]} \big]^{\T} \\
& \bm{\alpha}^{(t+1)} = \argmin_{\bm{\alpha}} \left\{ -\sum_{j = 1}^D \sum_{i = 1}^{n_j} \left[ \wt\omega_{ij}^{(t)}  \log\left(1 - h(\M{d}_{ij};\bm{\alpha}) \right) + (1 - \wt\omega_{ij}^{(t)}) \log\left(h(\M{d}_{ij};\bm{\alpha}) \right) \right] \right \},
\end{align*}
where $\wt{\omega}_{ij}^{(t)} = \frac{1}{U} \sum_{u = 1}^U (1 - m_{ij}^{[u]})$, $i=1,\ldots,n_j$, $j=1,\ldots,D$, and $\wt{\M{W}}^{(t)}$ denotes the corresponding diagonal matrix.   


\section{Calculations of the score and information matrix in \eqref{eq:covariance_louis}}\label{app:MSE_eblup}
We here derive explicit expressions for the terms $\nabla \mc{L}_j^{\textsf{c}}$ and 
$\nabla^2 \mc{L}_j^{\textsf{c}}$, $1 \leq j \leq D$. Given those expressions, the associated expectations and covariances on the right hand sides of \eqref{eq:covariance_louis} can then be approximated via Monte-Carlo integration using the sampling scheme in $\S$\ref{subsec:MCEM}. The specific approximations are provided as well and are indicated via the symbol $\cong$. We have 
{\small \begin{align*}
&\nabla_{\bm{\beta}} \mc{L}_{j}^{\textsf{c}}(\bm{\theta}) = \sum_{j = 1}^D \sum_{\ell_j \in \mc{P}_j} \mathbb{I}(\lambda_j = \ell_j)
\left( -\frac{\M{X}_{\ell_j}^{\T} (\M{y}_{\ell_j} - \M{X}_{\ell_j} \bm{\beta} - \M{Z}_{\ell_j} \bm{\gamma}_j)}{\sigma^2}  \right) \\
&\nabla_{\bm{\alpha}} \mc{L}_{j}^{\textsf{c}}(\bm{\theta}) = \sum_{j = 1}^D \sum_{\ell_j \in \mc{P}_j} \mathbb{I}(\lambda_j = \ell_j) \left( -\sum_{i \in \overline{\ell}_j} \frac{\nabla_{\bm{\alpha}} h(\M{d}_{ij};\bm{\alpha})}{h(\M{d}_{ij};\bm{\alpha})}  + \sum_{i \in \ell_j} \frac{\nabla_{\bm{\alpha}} h(\M{d}_{ij};\bm{\alpha})}{1 - h(\M{d}_{ij};\bm{\alpha})}  \right) \\
&\nabla_{\bm{\beta}\,\bm{\beta}}^2 \, \mc{L}_{j}^{\textsf{c}}(\bm{\theta}) = \sum_{j = 1}^D \sum_{\ell_j \in \mc{P}_j}  \mathbb{I}(\lambda_j = \ell_j) \frac{\M{X}_{\ell_j}^{\T} \M{X}_{\ell_j}}{\sigma^2}, \\ &\nabla_{\bm{\alpha}\,\bm{\alpha}}^2 \mc{L}_{j}^{\textsf{c}}(\bm{\theta}) = \sum_{j = 1}^D \sum_{\ell_j \in \mc{P}_j} \mathbb{I}(\lambda_j = \ell_j) \Bigg( \sum_{i \in \overline{\ell}_j} \frac{-h(\M{d}_{ij};\bm{\alpha}) \nabla_{\bm{\alpha} \bm{\alpha}}^2 h(\M{d}_{ij};\bm{\alpha}) + \nabla_{\bm{\alpha}} h(\M{d}_{ij};\bm{\alpha}) \{ \nabla_{\bm{\alpha}} h(\M{d}_{ij};\bm{\alpha}) \}^{\T}}{( h(\M{d}_{ij};\bm{\alpha}) )^2}  + \\
&\qquad \qquad \qquad \qquad \qquad \qquad \qquad + \sum_{i \in \ell_j}\frac{(1 - h(\M{d}_{ij};\bm{\alpha}))\nabla_{\bm{\alpha} \bm{\alpha}}^2 h(\M{d}_{ij};\bm{\alpha}) + \nabla_{\bm{\alpha}} h(\M{d}_{ij};\bm{\alpha}) \{ \nabla_{\bm{\alpha}} h(\M{d}_{ij};\bm{\alpha}) \}^{\T}}{(1 - h(\M{d}_{ij};\bm{\alpha}))^2}  \Bigg),
\end{align*}}
and finally $\nabla_{\bm{\beta}\,\bm{\alpha}}^2 = \M{0}$,  $1 \leq j \leq D$. Given the above expressions, we use the following Monte-Carlo approximations given 
samples $\{ \, \{ (m_{ij}^{[u]})_{i = 1}^{n_j}, \bm{\gamma}_j^{[u]} \}_{j = 1}^D \, \}_{u = 1}^U$ generated from the distributions of $\lambda_j, \bm{\gamma}_j | \mc{D}; \bm{\theta}$, $1 \leq j \leq D$: 
{\small \begin{align*}
&\E[\nabla_{\bm{\beta}} \mc{L}_{j}^{\textsf{c}}(\bm{\theta})|\mc{D}; \bm{\theta}] \cong 
-\frac{1}{U} \sum_{u = 1}^U \sum_{i = 1}^{n_j} (1 - m_{ij}^{[u]}) \M{x}_{ij} (y_{ij} - \M{x}_{ij}^{\T} \bm{\beta} - \M{z}_{ij}^{\T} \bm{\gamma}_j^{[u]})/\sigma^2, \\
&\E[\nabla_{\bm{\alpha}} \mc{L}_{j}^{\textsf{c}}(\bm{\theta})|\mc{D}; \bm{\theta}] \cong \frac{1}{U} \sum_{u = 1}^U \sum_{i = 1}^{n_j} \left(-m_{ij}^{[u]} \frac{\nabla_{\bm{\alpha}} h(\M{d}_{ij};\bm{\alpha})}{h(\M{d}_{ij};\bm{\alpha})} + (1 - m_{ij}^{[u]}) \frac{\nabla_{\bm{\alpha}} h(\M{d}_{ij};\bm{\alpha})}{1 - h(\M{d}_{ij};\bm{\alpha})} \right), \\
&\E[\nabla^2_{\bm{\beta} \, \bm{\beta}} \mc{L}_{j}^{\textsf{c}}(\bm{\theta})|\mc{D}; \bm{\theta}] \cong \frac{1}{U} \sum_{u = 1}^U \sum_{i = 1}^{n_j} (1 - m_{ij}^{[u]}) \M{x}_{ij} \M{x}_{ij}^{\T} / \sigma^2, \\
&\E[\nabla^2_{\bm{\alpha} \, \bm{\alpha}} \mc{L}_{j}^{\textsf{c}}(\bm{\theta})|\mc{D}; \bm{\theta}] \cong \frac{1}{U} \sum_{u = 1}^U \sum_{i = 1}^{n_j} \bigg( m_{ij}^{[u]} \frac{-h(\M{d}_{ij};\bm{\alpha}) \nabla_{\bm{\alpha} \bm{\alpha}}^2 h(\M{d}_{ij};\bm{\alpha}) + \nabla_{\bm{\alpha}} h(\M{d}_{ij};\bm{\alpha}) \{ \nabla_{\bm{\alpha}} h(\M{d}_{ij};\bm{\alpha}) \}^{\T}}{( h(\M{d}_{ij};\bm{\alpha}) )^2}  + \\
&\qquad \qquad \qquad \qquad \qquad \qquad \qquad (1 - m_{ij}^{[u]}) \frac{(1 - h(\M{d}_{ij};\bm{\alpha}))\nabla_{\bm{\alpha} \bm{\alpha}}^2 h(\M{d}_{ij};\bm{\alpha}) + \nabla_{\bm{\alpha}} h(\M{d}_{ij};\bm{\alpha}) \{ \nabla_{\bm{\alpha}} h(\M{d}_{ij};\bm{\alpha}) \}^{\T}}{(1 - h(\M{d}_{ij};\bm{\alpha}))^2} \bigg). 
\end{align*}}
Further note that $\nabla^2_{\bm{\beta} \bm{\alpha}} \mc{L}_j^{\textsf{c}} = \M{0}$. Moreover, we use 
{\small \begin{align*}
&\cov(\nabla_{\bm{\beta}} \mc{L}_{j}^{\textsf{c}}(\bm{\theta})|\mc{D}; \bm{\theta})
\cong \frac{1}{U} \sum_{u = 1}^U \sum_{i = 1}^{n_j} (1 - m_{ij}^{[u]}) \M{x}_{ij} \M{x}_{ij}^{\T} (y_{ij} - \M{x}_{ij}^{\T} \bm{\beta} - \M{z}_{ij}^{\T} \bm{\gamma}_j^{[u]})^2 / \sigma^4, \\
&\cov(\nabla_{\bm{\alpha}} \mc{L}_{j}^{\textsf{c}}(\bm{\theta})|\mc{D}; \bm{\theta}) \cong 
\frac{1}{U} \sum_{u = 1}^U \sum_{i = 1}^{n_j} \left(m_{ij}^{[u]} \frac{\nabla_{\bm{\alpha}} h(\M{d}_{ij};\bm{\alpha}) \nabla_{\bm{\alpha}} h(\M{d}_{ij};\bm{\alpha})^{\T}}{h(\M{d}_{ij};\bm{\alpha})^2} + (1 - m_{ij}^{[u]}) \frac{\nabla_{\bm{\alpha}} h(\M{d}_{ij};\bm{\alpha}) \nabla_{\bm{\alpha}} h(\M{d}_{ij};\bm{\alpha})^{\T}}{(1 - h(\M{d}_{ij};\bm{\alpha}))^2} \right), \\
&\cov(\nabla_{\bm{\beta}} \mc{L}_{j}^{\textsf{c}}(\bm{\theta}), \nabla_{\bm{\alpha}} \mc{L}_{j}^{\textsf{c}}(\bm{\theta}) | \mc{D}; \bm{\theta}) \cong \frac{1}{U} \sum_{u = 1}^U \sum_{i = 1}^{n_j} (1 - m_{ij}^{[u]}) \M{x}_{ij} \frac{(y_{ij} - \M{x}_{ij}^{\T} \bm{\beta} - \M{z}_{ij}^{\T} \bm{\gamma}_j^{[u]})}{\sigma^2} \frac{\nabla_{\bm{\alpha}} h(\M{d}_{ij};\bm{\alpha})^{\T}}{1 - h(\M{d}_{ij};\bm{\alpha})}.
\end{align*}}

\end{document}